\documentclass[pteplogo]{ptephy_v2}
\PassOptionsToPackage{dvipsnames}{xcolor}

\usepackage[
	colorlinks=true,
	pdfstartview=FitV,
	linkcolor=blue,
	citecolor=magenta,
	urlcolor=blue,
	bookmarks=true,
	bookmarksnumbered=true,
	pdftitle={},
	pdfauthor={}
]{hyperref}

\usepackage{bbm,bm,graphicx,mathtools,color,slashed}
\usepackage{amssymb,amsmath}
\usepackage[T1]{fontenc}

\usepackage[all]{xy}

\usepackage[framemethod=tikz]{mdframed}
\usepackage{tikz}

\usepackage[normalem]{ulem}  % \sout{old text} for strikeout
\usepackage{braket}

\usepackage{silence}
\newcommand{\cout}[1]{ \if 0 {#1} \fi }

\newcommand{\diff}{\mathrm{d}} 
\newcommand{\rmi}{\mathrm{i}} 
\newcommand{\rme}{\mathrm{e}}

\newcommand{\hH}{\hat{H}}
\newcommand{\hQ}{\hat{Q}}
\newcommand{\hG}{\hat{G}}

\newcommand{\tilAcal}{\widetilde{\mathcal{A}}}

\newcommand{\hK}{\hat{K}}
\newcommand{\hT}{\hat{T}}

\newcommand{\hU}{\hat{U}}

\newcommand{\ptc}[1]{{\bar{#1}}}

\newcommand{\hJ}{\hat{J}}

\newcommand{\ha}{\hat{a}}

\newcommand{\hP}{\hat{P}}

\newcommand{\hR}{\hat{R}}

\newcommand{\hPi}{\hat{\Pi}}
\newcommand{\hPsi}{\hat{\Psi}}

\newcommand{\hOcal}{\hat{\mathcal{O}}}

\newcommand{\hUcal}{\hat{\mathcal{U}}}

\newcommand{\pn}{\ptc{n}}

\newcommand\Lcal{\mathcal{L}}

\newcommand\Rcal{\mathcal{R}}
\newcommand\Acal{\mathcal{A}}

\newcommand\Rbb{\mathbb{R}}

\newcommand\SO{{\rm SO}}

\renewcommand\P{{\rm P}}

\newcommand{\ba}{\bm{a}}
\newcommand{\bb}{\bm{b}}
\newcommand{\bx}{\bm{x}}
\newcommand{\bX}{\bm{X}}
\newcommand{\by}{\bm{y}}
\newcommand{\bz}{\bm{z}}
\newcommand{\bk}{\bm{k}}
\newcommand{\bp}{\bm{p}}

\newcommand{\bnab}{\bm{\nabla}}
\newcommand{\bep}{\bm{\epsilon}}
\newcommand{\bzero}{\bm{0}}

\newcommand{\bA}{\bm{A}}
\newcommand{\bB}{\bm{B}}
\newcommand{\bD}{\bm{D}}

\newcommand{\bxi}{\bm{\xi}}

\newcommand{\ds}{\displaystyle}

\newcommand{\U}{\text{U}}

\newcommand{\with}{\quad\mathrm{with}\quad}

\tikzset{
  dotnum/.style={
    ringdot,
    minimum size=10pt,
    label=center:{\textcolor{black}{#1}}
  }
}

\begin{document}

\title{
Magnetic Symmetries and the Structure of Correlation Functions in Quantum Field Theory
}

\author[1,2,$\dag$]{Masaru Hongo}
\affil[1]{Department of Physics, Niigata University, Niigata 950-2181, Japan
% \email{hongo@phys.sc.niigata-u.ac.jp}
}
\affil[2]{RIKEN Center for Interdisciplinary Theoretical and Mathematical Sciences (iTHEMS), RIKEN, Wako 351-0198, Japan}

\author[3,$\ast$]{Kentaro Nishimura}
\affil[3]{Graduate School of Science and Technology, Niigata University, Niigata, 950-2181, Japan
% \email{nishiken.phys@gmail.com}
}

\affil[$\dag$]{\href{hongo@phys.sc.niigata-u.ac.jp}{hongo@phys.sc.niigata-u.ac.jp}}
\affil[$\ast$]{\href{nishiken.phys@gmail.com}{nishiken.phys@gmail.com}}

\begin{abstract}
 Quantum field theories in the presence of a static and uniform external magnetic field possess two characteristic spatial symmetries: magnetic translations and magnetic rotation.
 We investigate general consequences of these symmetries on correlation functions from a model-independent perspective, without relying on specific models or perturbative expansions.
 The projective structure of magnetic translation symmetry constrains correlation functions of charged operators to acquire the Schwinger phase and leads to a factorized form into a gauge-covariant phase factor and a reduced correlator depending only on relative coordinates.
 We further derive the spectral representation of two-point functions in terms of representations of the magnetic translation algebra, in which the Landau- and symmetric-gauge descriptions arise as different choices of basis.
 Our results provide a unified symmetry-based framework for quantum field theories in external magnetic fields.
\end{abstract}

\subjectindex{xxxx, xxx}

\maketitle

\section{Introduction}
\label{sec:intro}

Quantum systems under a magnetic field exhibit a variety of phenomena in atomic and condensed-matter physics, ranging from Zeeman splitting, Landau quantization, and magnetic phase transitions to the integer and fractional quantum Hall effects~\cite{Landau:1930,vonKlitzing:1980,Tsui:1982,Laughlin:1983}.
Moreover, quantum field theories (QFTs) in the presence of a strong magnetic field have also attracted considerable interest in high-energy physics~\cite{Heisenberg:1936nmg,Schwinger:1951nm}.
This is because magnetars in the universe and relativistic heavy-ion collision experiments at RHIC and the LHC can generate extremely strong magnetic fields, where QED plasmas and QCD matter are expected to exhibit nontrivial magnetic responses.
These systems provide a unique arena for studying quantum many-body phenomena under extreme conditions (see, e.g., Refs.~\cite{Miransky:2015ava,Andersen:2014xxa,Hattori:2023egw,Adhikari:2024bfa} for recent reviews on QFTs under magnetic fields).

The standard approach to quantum field theories in a magnetic field often relies on perturbative methods.
Typically, one starts from an explicit analysis of the corresponding free theory in a magnetic field, and then performs a perturbative expansion based on the two-point function exhibiting Landau quantization, or on equivalent formulations such as Schwinger's proper-time representation~\cite{Schwinger:1951nm}.
While this approach has the advantage of being explicitly tractable, it does not make it transparent whether the resulting properties persist beyond perturbative expansions, or are tied to specific approximations or models.

This issue becomes particularly evident in the structure of correlation functions.
Even in the presence of a spatially uniform and static magnetic field, the vector potential breaks ordinary translational invariance in a gauge-dependent manner, and the general form of correlation functions is not immediately transparent.
In practical calculations, it is well established that correlation functions acquire a position-dependent phase---the so-called Schwinger phase~\cite{Schwinger:1951nm}---which is necessary to ensure the proper gauge-covariant structure of correlation functions~(see Refs.~\cite{GomezDumm:2023owj,Miura:2025flc} for recent discussions).
However, the origin and necessity of this phase are often understood only at a technical level, without a clear identification of the underlying principle.
More generally, it remains unclear which features of correlation functions are robust and persist beyond specific computational schemes, and which depend on particular choices of gauge or approximation.

In this paper, we revisit quantum field theories in a magnetic field from a symmetry-based perspective and derive constraints on correlation functions on general grounds.
In particular, we focus on the genuine global symmetries of QFTs under a static and uniform magnetic field, namely magnetic translations~\cite{Brown1964,Zak1964-1,Zak1964-2} and a modified rotational symmetry around the magnetic-field direction, which we refer to as magnetic rotation.
We show that the Schwinger phase arises as a direct consequence of the projective realization of magnetic translation symmetry, leading to a factorized form of correlation functions into a gauge-covariant phase factor and a reduced correlator depending only on relative coordinates.
We further derive a generalization of the Umezawa--Kamefuchi--K\"all\'en--Lehmann spectral representation~\cite{Umezawa:1951rp,Kallen:1952zz,Lehmann:1954xi} organized by representations of the magnetic translation algebra.
These structures are dictated solely by magnetic translation and rotation symmetries and therefore hold independently of microscopic details or perturbative approximations.

The organization of this paper is as follows.
In Sec.~\ref{sec:symmetry}, we formulate the magnetic translation and magnetic rotation symmetries of QFTs in a uniform magnetic field on general grounds, derive the associated symmetry algebra, and discuss its gauge covariance.
In Sec.~\ref{sec:correlation-function}, we derive general constraints on correlation functions, including the selection rules and the Schwinger-phase factorization implied by magnetic translation symmetry, and extend these results to higher-point functions.
In Sec.~\ref{sec:spectral}, we derive a general spectral representation of two-point functions organized by representations of the magnetic translation algebra.
Finally, we conclude in Sec.~\ref{sec:summary} with a summary and outlook.

\section{Magnetic symmetries of QFT in a uniform magnetic field}
\label{sec:symmetry}

Quantum field theories in a uniform external magnetic field possess characteristic spatial symmetries: magnetic translations in the transverse plane and the magnetic rotation around the direction of the magnetic field.
In this section, we formulate these symmetries in a general and model-independent manner.
After defining their action on dynamical fields in Sec.~\ref{sec:mag-translation-rotation}, we derive the Noether charges and symmetry algebra in Sec.~\ref{sec:generator-algebra}.
In Sec.~\ref{sec:symmetry-operator}, we introduce the symmetry operators and classify local operators by their quantum numbers.
We finally clarify the gauge covariance of the magnetic symmetries in Sec.~\ref{sec:gauge-covariance}.

\subsection{Magnetic translation and magnetic rotation}
\label{sec:mag-translation-rotation}

We consider a $(d+1)$-dimensional system consisting of both charged and neutral particles, described by a set of fields $\Psi^n$ with electric charge $q_n$ and spin $s_n$.
In the absence of a background magnetic field, the system is invariant under spacetime translations, a global $\U(1)$ transformation, and spatial rotations $\SO(d)$.
We do not assume boost symmetry in general, and the following discussion applies equally to relativistic and nonrelativistic theories.
These symmetries act on the fields $\Psi^n$ as
\begin{subequations}\label{eq:symmetry-without-B}
\begin{align}
 \Rbb_t:~\Psi^n(t,\bx) &\to \Psi^n(t+\epsilon, \bx),
 \\
 \U(1)_{\mathrm{global}}:~\Psi^n(t,\bx) &\to \rme^{\rmi q_n \alpha} \Psi^n (t,\bx), 
 \label{eq:U(1)-transformation}
 \\
 \Rbb_s^d:~\Psi^n(t,\bx) &\to \Psi^n(t, \bx + \ba),
 \\
 \SO(d):~\Psi^n(t,\bx) &\to U^{(s_n)}_R \Psi^n(t, R \bx),
\end{align}
\end{subequations}
where $U_{R}^{(s_n)}$ denotes the spin-$s_n$ representation matrix.

We now introduce a uniform background magnetic field along the $z$-axis (applicable to both $d=2$ and $d=3$).
This is implemented by introducing a background $\U(1)$ gauge field $A_\mu$ whose field strength is given by
\begin{equation}
 F_{\mu\nu} := \partial_\mu A_\nu - \partial_\nu A_\mu
 = (\delta_\mu^1 \delta_\nu^2 - \delta_\nu^1 \delta_\mu^2) B,
\end{equation}
where $B$ is a time-independent constant specifying the strength of the uniform magnetic field.

We then assume that the coupling of the fields $\Psi^n$ to the electromagnetic field consists of two types.
The first is the minimal coupling described by the covariant derivative
\begin{equation}
 D_\mu \Psi^n := \partial_\mu \Psi^n - \rmi q_n A_\mu \Psi^n,
 \label{eq:covariant-derivative}
\end{equation}
which ensures covariance under local $\U(1)$ transformations
\begin{equation}
 \U(1)_\mathrm{local}:~
 \begin{cases}
  \Psi^n(t,\bx) \to \rme^{\rmi q_n \alpha (t,\bx)} \Psi^n (t,\bx),
  \\
  A_\mu (t,\bx) \to A_\mu + \partial_\mu \alpha(t,\bx).
 \end{cases}
 \label{eq:gauge-tr}
\end{equation}
The second type of coupling corresponds to non-minimal interactions with the background magnetic field, directly involving the field strength, with $F_{12} = -F_{21} = B$.

The dynamics of the system is then described by an action
\begin{equation}
 S[\Psi^n; A_\mu]
 = \int \diff t \diff^d x \, \mathcal{L}(\Psi^n, D_\mu \Psi^n, F_{\mu\nu}),
 \label{eq:general-action}
\end{equation}
where $\mathcal{L}$ is a local functional of the fields, their covariant derivatives, and the background field strength.
We do not assume a specific form of $\mathcal{L}$, except that it is invariant under the symmetries in Eq.~\eqref{eq:symmetry-without-B} in the absence of the background magnetic field.

Let us specify the \textit{global symmetry} of the present setup; namely, a general quantum field theory in a uniform background magnetic field.
First of all, the time-translation symmetry $\Rbb_t$, the global $\U(1)$ symmetry, and the spatial translation along the $z$-direction $\Rbb_z$ remain unbroken in the presence of a background magnetic field.
On the other hand, the magnetic field selects the $z$-direction and explicitly breaks the rotational symmetry $\SO(d)$ down to its subgroup $\SO(2)_z$.

The fate of the remaining spatial symmetries, namely the transverse translations $\Rbb_x \times \Rbb_y$ and the rotation around the $z$-axis $\SO(2)_z$, requires a more careful analysis.
Although the magnetic field is uniform and invariant under rotations around the $z$-axis, the minimal coupling is not determined solely by the gauge-invariant field strength $F_{12}=-F_{21}=B$, but is implemented through the covariant derivative \eqref{eq:covariant-derivative}.
In particular, the transverse components of the background gauge potential, $A_{i_\perp} = (A_1,A_2)$, depend explicitly on the transverse coordinates $x^i_\perp := (x^1,x^2)$.
Accordingly, the transverse translations and the rotation around the $z$-axis are replaced by new symmetry transformations: the former correspond to the well-known \textit{magnetic translations}, while the latter gives rise to a \textit{magnetic rotation}.

\paragraph{a) Magnetic translation}
Let us begin with the magnetic translation~\cite{Brown1964,Zak1964-1,Zak1964-2}.
In the presence of a background magnetic field, the gauge field depends explicitly on the spatial coordinates, and ordinary translational invariance is lost.
Nevertheless, spatial translations supplemented by a position-dependent $\U(1)$ phase transformation preserve the form of the covariant derivative.

To see this, we consider the translation in the transverse plane:
\begin{equation}
 \Psi^n (t,\bx)
 \to 
 \Psi^{n\prime} (t,\bx)
 = \rme^{-\rmi q_n f (\bx_\perp+\ba_\perp,\bx_\perp)} 
 \Psi^n (t,\bx + \ba_\perp),
 \label{eq:magnetic-translation}
\end{equation}
where the phase function $f (\bx_\perp+\ba_\perp,\bx_\perp)$ associated with a translation by $\ba$ is required to satisfy
\begin{equation}
 \partial_i f (\bx_\perp+\ba_\perp,\bx_\perp)
 =  A_i (\bx_\perp + \ba_\perp) - A_i (\bx_\perp) .
 \label{eq:condition-magnetic-translation}
\end{equation}
Such a function exists because the right-hand side is curl-free:
\begin{equation}
 \bnab \times [\bA (\bx_\perp + \ba_{\perp}) - \bA (\bx_\perp) ] = 
 \bB (\bx_\perp + \ba_\perp) - \bB(\bx_\perp) = 0,
 \label{eq:condition-magnetic-translation-rot}
\end{equation}
where we used the assumption that the magnetic field is uniform.
Therefore, integrating Eq.~\eqref{eq:condition-magnetic-translation} along an arbitrary path $C$ connecting a reference point $\bx_{0\perp}$ to $\bx_\perp$, we obtain
\begin{equation}
 f (\bx_\perp+\ba_\perp,\bx_\perp)
 = \int_C
 \diff \xi^i 
 \big[ A_i(\bxi+ \ba_\perp) - A_i(\bxi) \big]
 + f (\bx_{0\perp}+\ba_\perp,\bx_{0\perp}) .
 \label{eq:finite-f}
\end{equation}
In the case of the infinitesimal translation, expanding both sides with respect to $\ba$, we find 
\begin{align}
 &a_\perp^i
 \left[
  \frac{\partial f (\bx_\perp+\ba_\perp,\bx_\perp)}{\partial a_\perp^i}
  - \frac{\partial f (\bx_{0\perp}+\ba_\perp,\bx_{0\perp})}{\partial a_\perp^i}
 \right]_{\ba_{\perp} = 0}
 \notag \\
 =& \int_{C}
 \diff \xi^i a_\perp^j \partial_j A_i(\bxi)
 = a_\perp^j \int_{C}
 \diff \xi^i 
 \big[
  \partial_i A_j(\bxi) + F_{ji} 
 \big]
 \notag \\
 =& a_\perp^i 
 \left[
  A_i (\bx_{\perp}) - A_i (\bx_{0\perp})
 + F_{ij} (x^j_\perp - x_{0\perp}^j)
 \right] ,
\end{align}
where we used $f (\bx_\perp,\bx_\perp) = 0$.
This equation implies
\begin{equation}
 \left.
 \frac{\partial f (\bx_\perp+\ba_\perp,\bx_\perp)}{\partial a^i}
 \right|_{\ba_\perp = 0}
 = A_i (\bx_{\perp}) + F_{ij} x^j_{\perp}.
 \label{eq:f-a-derivative}
\end{equation}
Here, we have set the $\bx_\perp$-independent constant of $f (\bx_\perp+\ba_\perp,\bx_\perp)$ to zero, since it only contributes to an overall phase of the magnetic translation operator.

The crucial point here is that under Eq.~\eqref{eq:magnetic-translation} the covariant derivative transforms as
\begin{align}
 D_i \Psi^n (t,\bx)
 &\to 
 [\partial_i - \rmi q_n A_i (\bx_\perp) ]
 \rme^{- \rmi q_n f (\bx_\perp+\ba_\perp,\bx_\perp)}
 \Psi^n (t,\bx + \ba_\perp)
 \notag \\
 &= \rme^{- \rmi q_n f (\bx_\perp+\ba_\perp,\bx_\perp)}
 [\partial_i - \rmi q_n 
 \{A_i (\bx_\perp) + \partial_i f (\bx_\perp+\ba_\perp,\bx_\perp) \}]
 \Psi^n (t,\bx +\ba_\perp)
 \notag \\
 % &= \rme^{- \rmi q_n f (\bx_\perp+\ba_\perp,\bx_\perp)}
 % [\partial_i - \rmi q_n A_i (\bx_\perp + \ba) ]
 % \Psi^n (t,\bx + \ba_\perp)
 % \notag \\
 &= \rme^{-\rmi q_n f (\bx_\perp+\ba_\perp,\bx_\perp)}
 D_i \Psi^n (t,\bx+\ba_\perp),
\end{align}
where we used Eq.~\eqref{eq:condition-magnetic-translation} in the second line.
This transformation rule shows that the covariant derivative transforms covariantly under magnetic translations.
As a result, if the original theory is invariant under Eq.~\eqref{eq:symmetry-without-B}, the action \eqref{eq:general-action} in the presence of a uniform magnetic field is invariant under the transformation in Eqs.~\eqref{eq:magnetic-translation}-\eqref{eq:condition-magnetic-translation} as $S[\Psi^{n\prime}; A_\mu] = S[\Psi^n; A_\mu]$.
This defines the magnetic translation symmetry.

\paragraph{b) Magnetic rotation}
We next consider the magnetic rotation symmetry.

In analogy with magnetic translations, we introduce a combination of a spatial rotation $R_\theta \in \SO(2)_z$ with rotation angle $\theta$ and a position-dependent $\U(1)$ phase transformation:
\begin{equation}
 \Psi^n (t,\bx)
 \to 
 \Psi^{n\prime} (t,\bx)
 = \rme^{-\rmi q_n g_\theta(\bx_\perp)} 
 U_{\theta}^{(s_n)}
 \Psi^n (t, R_\theta \bx),
 \label{eq:magnetic-rotation}
\end{equation}
where $U_{\theta}^{(s_n)} = \rme^{\rmi \theta (S_z)_n}$ is the spin representation matrix defined through the generator $(S_z)_n$ of $\SO(2)_z$ acting on $\Psi^n$.
Here, the phase function $g_\theta(\bx_\perp)$ associated with a $\SO(2)_z$ rotation by angle $\theta$ is required to satisfy
\begin{equation}
 \partial_i g_\theta(\bx_\perp)
 = (R_\theta)_i^{~j} A_j (R_\theta \bx_\perp) -  A_i (\bx_\perp).
 \label{eq:condition-magnetic-rotation}
\end{equation}
Such a function exists locally because the right-hand side is again curl-free:%
% \footnote{
% [Note]
% Defining $\bX_\perp := R_\theta^{-1} \bx_\perp$, we can explicitly calculate
% \begin{equation}
%  \begin{split}
%   [\bnab_x \times R_\theta^{-1} \bA (R_\theta^{-1} \bx_{\perp})]^i
%   &= \epsilon^{ijk} 
%   \partial_j^x (R_\theta^{-1})_k^{~l} 
%   A_l (\underbrace{R_\theta^{-1} \bx_{\perp}}_{=\bX})
%   = \epsilon^{ijk} (R_\theta^{-1})_k^{~l} 
%   (\partial_j^x X^m) 
%   \partial_m^X A_l (\bX_{\perp})
%   \\
%   &= 
%   \underbrace{
%   \epsilon^{ijk} (R_\theta^{-1})_k^{~l} (R_\theta^{-1})^m_j 
%   }_{=(R_\theta)^i_{~a} \epsilon^{aml} }
%   \partial_m^X A_l (\bX_{\perp})
%   = [R_\theta \bnab \times \bA (\bX_{\perp})]^i
%   = [R_\theta \bB (\bX_{\perp})]^i
%   \\
%  \end{split}
% \end{equation}
% }:
\begin{equation}
 \bnab \times 
 \left[
  R_\theta \bA (R_\theta \bx_\perp) - \bA (\bx_\perp) 
 \right]
 = R_\theta^{-1} \bB (R_\theta \bx_\perp) - \bB(\bx_\perp)  = 0,
 \label{eq:condition-magnetic-rotation-curl}
\end{equation}
where we have used the assumption that the magnetic field is $\SO(2)_z$-invariant.
Integrating Eq.~\eqref{eq:condition-magnetic-rotation} along an arbitrary path connecting a reference point $\bx_{0\perp}$ to $\bx_\perp$, we obtain 
\begin{equation}
 g_\theta(\bx_\perp)
 = \int_{C}
 \diff \xi^i 
  \big[ 
  (R_\theta)_i^{~j} A_j (R_\theta \bxi) -  A_i(\bxi) 
  \big]
  + g_\theta (\bx_{0\perp}) .
 \label{eq:finite-g}
\end{equation}
Expanding this equation with respect to $\theta$, we obtain
\begin{align}
 \theta
 \left[
  \frac{\partial g_{\theta}(\bx_\perp)}{\partial \theta}
  - \frac{\partial g_{\theta}(\bx_{0\perp})}{\partial \theta}
 \right]_{\theta = 0}
 &= \theta \int_{C}
 \diff \xi^i 
 \left[
  \epsilon_i^{~j} A_j (\bxi)
  + \epsilon^j_{~k} \xi^k \partial_j A_i (\bxi)
 \right]
 \notag \\
%  &= \theta \int_{C}
%  \diff \xi^i 
%  \left[
%   \epsilon_i^{~j} A_j (\bxi)
%   + \epsilon^j_{~k} \xi^k (F_{ji} + \partial_i A_j)
%  \right]
%  \notag \\
 &= \theta \int_{C}
 \diff \xi^i 
 \left[
  \partial_i (\epsilon^j_{~k} \xi^k A_j)
  + \epsilon^j_{~k} \xi^k F_{ji}
 \right]
 \notag \\
 &= \theta 
 \left[
  \hspace{-1pt}
  \Big(
    \bx_{\perp} \times \bA (\bx_{\perp}) 
    - \bx_{0\perp} \times \bA (\bx_{0\perp}) 
  \Big)_z
  \hspace{-6pt}
  - \frac{B}{2} (\bx_{\perp}^2 - \bx_{0\perp}^2 )
 \right] ,
\end{align}
where we used $g_{0} (\bx_{\perp}) = 0$.
This equation determines the derivative of $g_\theta (\bx_{\perp})$ up to an $\bx_\perp$-independent constant. 
Setting this constant to zero, we obtain
\begin{equation}
 \left.
 \frac{\partial g_{\theta}(\bx_\perp)}{\partial \theta}
 \right|_{\theta = 0}
 = \big[ 
    \bx_{\perp} \times \bA (\bx_{\perp}) 
   \big]_z
   - \frac{B}{2} \bx_{\perp}^2.
 \label{eq:g-theta-derivative}
\end{equation}

Then, we can show that under Eq.~\eqref{eq:magnetic-rotation} the covariant derivative transforms as
\begin{equation}
 \begin{split}
 D_i \Psi^n (t,\bx)
 &\to 
 [\partial_i^x - \rmi q_n A_i (\bx_\perp) ]
 \rme^{-\rmi q_n g_\theta(\bx_\perp)}
 U_{\theta}^{(s_n)} \Psi^n (t, R_\theta\bx)
 \\
 &= \rme^{- \rmi q_n g_\theta(\bx_\perp)} U_{\theta}^{(s_n)}
 [\partial_i^x - \rmi q_n 
 \{A_i (\bx_\perp) + \partial_i g_\theta(\bx_\perp) \}]
 \Psi^n (t, R_\theta \bx)
 \\
 % &= \rme^{-\rmi q_n g_\theta(\bx_\perp)}
 % U_{\theta}^{(s_n)} (R_\theta)_i^{~j}
 % [\partial_j^{X} - \rmi q_n A_j (\bX_{\perp}) ]
 % \Psi^n (t,\bX_\perp)
 % \\
 &= \rme^{-\rmi q_n g_\theta(\bx_\perp)} 
 U_{\theta}^{(s_n)} (R_\theta)_i^{~j}
 D_j \Psi^n (t,\bX),
 \end{split}
\end{equation}
where we introduced $\bX := R_\theta \bx$ and used Eq.~\eqref{eq:condition-magnetic-rotation} in the second line.
This transformation rule shows that the covariant derivative transforms covariantly under the magnetic rotation.
As a result, if the original theory is invariant under Eq.~\eqref{eq:symmetry-without-B}, the action \eqref{eq:general-action} in the presence of a uniform magnetic field is also invariant 
under the transformation in Eqs.~\eqref{eq:magnetic-rotation}-\eqref{eq:condition-magnetic-rotation} as $S[\Psi^{n\prime}; A_\mu] = S[\Psi^n; A_\mu]$.
This defines the magnetic rotation symmetry.

\subsection{Noether charges and magnetic symmetry algebra}
\label{sec:generator-algebra}

We now derive the Noether charges and the resulting symmetry algebra associated with the magnetic symmetries.

Let us first derive the infinitesimal forms of magnetic translations and the magnetic rotation.
Expanding Eqs.~\eqref{eq:finite-f} and \eqref{eq:finite-g}, we find 
\begin{align}
 f (\bx_\perp+\ba_\perp,\bx_\perp)
 &= a_\perp^i 
 \Big[
  A_i (\bx_\perp) + F_{ij} x_\perp^j
 \Big] 
 + O(\ba_\perp^2)
 ,
 \\
 g_\theta(\bx_\perp)
 &=
 \theta 
 \left[
  \big(
   \bx_{\perp} \times \bA (\bx_\perp) 
  \big)_z
  % \epsilon^i_{~j} 
  % \big\{ x^j A_i (\bx_\perp) - x_0^j A_i (\bx_{0\perp}) \big\}
   - \frac{B}{2} \bx_\perp^2 
 \right]
 + O(\theta^2)
  ,
\end{align}
where we used Eqs.~\eqref{eq:f-a-derivative} and \eqref{eq:g-theta-derivative}.
We then find that the infinitesimal magnetic translations and rotation act on the dynamical fields as
\begin{align}
 \delta^{\mathrm{tr}}_i \Psi^n (t,\bx)
 &= \partial_i \Psi^n (t,\bx)
 - \rmi q_n \big[ A_i (\bx_\perp) + F_{ij} x_\perp^j \big] 
 \Psi^n (t,\bx)
 \notag \\
 &= D_i \Psi^n (t,\bx) - \rmi q_n F_{ij} x_\perp^j \Psi^n (t,\bx),
 \\
 \delta^{\mathrm{rot}} \Psi^n (t,\bx)
 &= \big[
     (\bx_{\perp} \times \bnab )_z + 
     \rmi (S_z)_n
      \big] \Psi^n (t,\bx)
  - \rmi q_n 
  \left[
   \big( \bx_\perp \times \bA (\bx_\perp) \big)_z - \frac{B}{2} \bx_\perp^2
  \right] 
 \Psi^n (t,\bx)
 \notag \\
 &= \rmi \big[
     (\bx_{\perp} \times \rmi^{-1} \bD )_z + (S_z)_n
      \big] \Psi^n (t,\bx)
  + \frac{\rmi q_n B}{2} \bx_\perp^2 \Psi^n (t,\bx),
\end{align}
where $(S_z)_n$ denotes the generator of $\SO(2)_z$ rotations in the spin-$s_n$ representation.

Using these infinitesimal transformations, we define the Noether charges associated with magnetic translations and the magnetic rotation as 
\begin{align}
 \hK_i 
 &:= - \int \diff^d x 
 \left[
 \hPi_n D_i \hPsi^n 
 - \rmi q_n F_{ij} x_\perp^j \hPi_n \hPsi^n 
 \right]
 \qquad 
 (i=1,2),
 \label{eq:def-magnetic-momentum}
 \\
 \hJ_z 
 &:= -\int \diff^d x 
 \left[
  \rmi 
  \hPi_n 
  \big[
   (\bx_{\perp} \times \rmi^{-1} \bD )_z + (S_z)_n
    \big] \hPsi^n 
   + \frac{\rmi q_n B}{2} \bx_\perp^2 \hPi_n \hPsi^n 
  \right].
 \label{eq:def-magnetic-angular-momentum}
\end{align}
Here, we define the fields conjugate to the $\Psi^n$ by 
$\Pi_n (t,\bx) := \partial \Lcal/\partial (\partial_t \Psi^n)$, and promote $\Psi^n$ and $\Pi_n$ to operators satisfying the canonical quantization conditions
\begin{equation}
 \big[\hPsi^n (t,\bx), \hPi_{\pn} (t,\by) \big]_{\mp} 
 = \rmi \delta^n_{\pn} \delta (\bx - \by),
\end{equation}
where $\mp$ denotes the commutator or anti-commutator depending on the quantum statistics of the field $\hPsi^n$.
For completeness, we also introduce the Noether charges attached to the global $\U(1)$ symmetry, the longitudinal translation $\Rbb_z$, and time translation as
\begin{align}
 \hQ &:=
 - \int \diff^d x \,
 \rmi q_n \hPi_n \hPsi^n,
 \\
 \hP_z &:=
 - \int \diff^d x \,
 \hPi_n \partial_z \hPsi^n,
 \\
 \hH &:=
 \int \diff^d x 
 \big[
 \hPi_n \partial_t \hPsi^n - \Lcal 
 \big].
\end{align}

It is worth noting that $\hK_i$ differs from the generator of ordinary spatial translations in the absence of a magnetic field.
In particular, $\hK_i$ involves both the covariant derivative and the magnetic-field-dependent term, and should be distinguished from the canonical momentum.
This distinction reflects the fact that, in the presence of a background magnetic field, the generator of physical translations is modified to include the background gauge field, and the canonical momentum no longer generates a symmetry.

Using the Noether charges and the canonical commutation relations, we obtain the following symmetry algebra for magnetic translations and the magnetic rotation:
\begin{align}
 \big[\hK_i,\hK_j\big] &= - \rmi F_{ij} \hQ \qquad (i,j=1,2) ,
 \label{eq:KK-commutator}
 \\
 \big[\hJ_z,\hK_i\big] &= \rmi \epsilon_{ij} \hK_j ,
 \label{eq:JK-commutator}
\end{align}
with the anti-symmetric tensor $\epsilon_{12} = 1 = - \epsilon_{21}$.
The remaining commutators involving $\hQ$ and $\hH$ all vanish.
Note that Eq. \eqref{eq:KK-commutator} demonstrates the non-Abelian nature of the magnetic translation symmetry.
This shows that the magnetic translation algebra exhibits a central extension with $\hQ$ as the center element.

\subsection{Symmetry operators and operator classification}
\label{sec:symmetry-operator}

Based on the Noether charges, we introduce the symmetry operators associated with the $\U(1)$ symmetry and the magnetic translation and rotation symmetries
\begin{align}
 \hU(\alpha) &:= \rme^{\rmi \alpha \hQ},
 \\
 \hT(\ba) &:= \rme^{\rmi a_\perp^i \hK_i + \rmi a^z \hP_z}
 = \rme^{-\frac{\rmi}{2}B a_x a_y \hQ}
 \rme^{\rmi a_x\hK_x} \rme^{\rmi a_y\hK_y} \rme^{\rmi a_z\hP_z},
 \label{eq:magnetic-translation-operator}
 \\
 \hR(\theta) &:= \rme^{\rmi \theta \hJ_z}.
\end{align}
where the second expression of $\hT(\ba)$ follows from the Baker--Campbell--Hausdorff formula.
Moreover, using the commutation relation \eqref{eq:KK-commutator}, we find
\begin{equation}
 \hT(\ba) \hT(\bb)
 = \rme^{\frac{\rmi}{2} F_{ij} a_\perp^i b_\perp^j\, \hQ}
 \hT(\ba+\bb),
 \label{eq:projective-rep.}
\end{equation}
which shows that the magnetic translation operators form a projective representation of the translation group.%
\footnote{
For group elements $a,b$, a projective representation $\Rcal$ is defined by
\begin{equation}
 \Rcal(a) \Rcal(b) = \rme^{\rmi \phi (a,b)} \Rcal(a+b),
\end{equation}
where the nontrivial phase factor $\phi(a,b)$ satisfies the associativity condition (or 2-cocycle condition)
\begin{equation}
 \phi (a,b) + \phi (a+b,c) 
 = \phi (a,b+c) + \phi (b,c).
\end{equation}
}
We further introduce the spacetime translation operator 
\begin{equation}
 \hUcal (t,\bx)
 := \rme^{- \rmi \hH t + \rmi \hP_z z + \rmi \hK_i x_{\perp}^i}
 = \rme^{ - \rmi \hH t + \rmi \hP_z z }
 \hT(\bx_{\perp}),
 \label{eq:translation-op-decomposition}
\end{equation}
which satisfies 
\begin{equation}
 \hUcal (t_1,\bx_1) \hUcal (t_2,\bx_2) = 
 \rme^{+ \frac{\rmi}{2} F_{ij} x_{1\perp}^i x_{2\perp}^j\, \hQ}
 \hUcal (t_1+t_2,\bx_1+\bx_2).
 \label{eq:projective-spacetime-translation}
\end{equation}

These symmetry operators act on the elementary operators $\hPsi^n$ as
\begin{align}
 \hU^\dag (\alpha) \hPsi^n(t,\bx) \hU (\alpha)
 &=
 \rme^{\rmi q_n \alpha}
 \hPsi^n(t,\bx),
 \label{eq:phase_quantum}
 \\
 \hT^\dag(\ba) \hPsi^n(t,\bx) \hT(\ba)
 &=
 \rme^{-\rmi q_n f (\bx_\perp+\ba_\perp,\bx_\perp)}
 \hPsi^n(t,\bx+\ba),
 \label{eq:magnetic_translation_quantum}
 \\
 \hR^\dag(\theta) \hPsi^n(t,\bx) \hR (\theta)
 &=
 \rme^{-\rmi q_n g_\theta(\bx_\perp)}
 U_{\theta}^{(s_n)}
 \hPsi^n(t,R_\theta \bx),
 \label{eq:magnetic_rotation_quantum}
\end{align}
which reproduce the transformations given in Eqs.~\eqref{eq:U(1)-transformation}, \eqref{eq:magnetic-translation} and \eqref{eq:magnetic-rotation}, respectively.
Moreover, consistency with the projective composition law \eqref{eq:projective-rep.} requires the phase functions to satisfy
\begin{equation}
 f (\bz_\perp,\by_\perp) + f (\by_\perp,\bx_\perp)
 = f (\bz_\perp,\bx_\perp)
 + \frac{1}{2} F_{ij} (z_\perp - y_\perp)^i (y_\perp - x_\perp)^j ,
 \label{eq:composition-rule-f}
\end{equation}
which has the structure of a 2-cocycle condition reflecting the projective nature of the magnetic translation symmetry \eqref{eq:projective-rep.}.
Similarly, consistency of finite magnetic rotations requires
\begin{equation}
 g_\varphi(R_\theta \bx_\perp)
 +
 g_\theta(\bx_\perp)
 =
 g_{\theta+\varphi}(\bx_\perp).
 \label{eq:composition-rule-g}
\end{equation}
These relations constrain the phase functions 
$f(\by_\perp,\bx_\perp)$ and $g_\theta (\bx_{\perp})$.
The cocycle structure in Eq.~\eqref{eq:composition-rule-f} will play a central role in determining the Schwinger-type phase of correlation functions in Sec.~\ref{sec:schwinger-phase}.

We now classify local operators according to their quantum numbers under the magnetic symmetry algebra.
Let $\hOcal_a(x)$ be a local operator carrying the $\U(1)$ charge $q_a$ and the intrinsic spin $s_a$ under $\SO(2)_z$.
These quantum numbers are defined by the commutation relations at the origin, where the orbital contribution vanishes:
\begin{align}
 \big[ \hQ,\hOcal_a(0) \big]
 &= - q_a \hOcal_a(0),
 \label{eq:def-charge}
 \\
 \big[ \hJ_z,\hOcal_a(0) \big]
 &= - s_a \hOcal_a(0).
 \label{eq:def-spin}
\end{align}
Any local operator $\hOcal_a(t,\bx)$ transforms under the $\U(1)$ symmetry, magnetic translations, and magnetic rotations as
\begin{align}
 \hU^\dag (\alpha) \hOcal_a(t,\bx) \hU (\alpha)
 &=
 \rme^{\rmi q_a \alpha}
 \hOcal_a(t,\bx),
 \label{eq:operator-phase-transformation}
 \\
 \hT^\dag (\ba) \hOcal_a(t,\bx) \hT (\ba)
 &=
 \rme^{-\rmi q_a f (\bx_\perp+\ba_\perp,\bx_\perp)}
 \hOcal_a(t,\bx+\ba),
 \label{eq:operator-magnetic-translation}
 \\
 \hR^\dag(\theta) \hOcal_a(t,\bx) \hR(\theta)
 &=
 \rme^{-\rmi q_a g_\theta(\bx_\perp)}
 \rme^{\rmi s_a \theta}
 \hOcal_a(t,R_\theta\bx).
 \label{eq:operator-magnetic-rotation}
\end{align}
Combining the above transformations, the spacetime translation acts on $\hOcal_a(t,\bx)$ as
\begin{equation}
 \hUcal^\dag (\epsilon) \hOcal_a(x) \hUcal (\epsilon)
 = \rme^{-\rmi q_a f (\bx_\perp+\bep_\perp,\bx_\perp)} 
 \hOcal_a(x+\epsilon),
 \label{eq:spacetime-translation}
\end{equation}
where we used $x = (t,\bx) = (t,\bx_\perp,z)$ and $\epsilon = (\epsilon^0,\bep) = (\epsilon^0,\bep_\perp,\epsilon^z)$.

\subsection{Gauge covariance of magnetic symmetries}
\label{sec:gauge-covariance}

We finally discuss how the magnetic symmetries are compatible with the local $\U(1)$ gauge transformation in Eq.~\eqref{eq:gauge-tr}.

First of all, it is worth emphasizing that the magnetic translation and rotation symmetries defined by Eqs.~\eqref{eq:magnetic-translation}-\eqref{eq:condition-magnetic-translation}
and Eqs.~\eqref{eq:magnetic-rotation}-\eqref{eq:condition-magnetic-rotation} act only on the dynamical fields $\hPsi^n$ and not on the background gauge field $A_\mu$.
This shows that the magnetic symmetries are realized as genuine global symmetries of the system, rather than as a gauge redundancy associated with Eq.~\eqref{eq:gauge-tr}.
Nevertheless, their explicit forms involve the compensating phase functions $f (\bx_\perp+\ba_\perp,\bx_\perp)$ and $g_\theta(\bx_\perp)$, which themselves depend on the gauge choice for the background field.
We now determine how these phase functions transform under the gauge transformation~\eqref{eq:gauge-tr}.

To investigate the gauge covariance, we explicitly regard the magnetic translation operator as a functional of both the dynamical fields and the background gauge field as $\hT (\ba) = \hT_{\ba} [\hPsi^n;A_\mu]$.
The crucial point is that this operator is gauge invariant [recall Eq.~\eqref{eq:def-magnetic-momentum}], so that it satisfies 
\begin{equation}
 \hT_{\ba} [\rme^{\rmi q_n \alpha} \hPsi^n; A_\mu + \partial_\mu \alpha]
 = \hT_{\ba} [\hPsi^n; A_\mu]. 
\end{equation}
The local $\U(1)$ transformation acting on the dynamical fields is generated by the operator
\begin{equation}
 \hG [\alpha] := 
 \exp \left( \rmi \int \diff^d x \, \alpha (x) \hJ^0 (x)\right)
 \with 
 \hJ^\mu (x) = \frac{\delta S}{\delta A_\mu (x)},
\end{equation}
where $\hJ^\mu(x)$ denotes the gauge current.
Since $\hG[\alpha]$ acts only on the dynamical fields and not on the background gauge field, we have
\begin{equation}
 \hG [\alpha] \hT_{\ba}[\hPsi^n; A_\mu] \hG^\dag [\alpha] = 
 \hT_{\ba}[\rme^{-\rmi q_n \alpha} \hPsi^n; A_\mu].
\end{equation}

Let us now consider the magnetic translation of the operator $\hPsi^n$ in Eq.~\eqref{eq:magnetic_translation_quantum}.
Multiplying Eq.~\eqref{eq:magnetic_translation_quantum} by $\hG$ from the left and by $\hG^\dag$ from the right, and inserting the identity operator $\bm{1} = \hG^\dag \hG$, we find
\begingroup
\fontsize{10.4}{11.5}\selectfont
\begin{align}
 \rme^{-\rmi q_n f_A (\bx_\perp+\ba_\perp,\bx_\perp)}
 \rme^{-\rmi q_n \alpha (x+\ba)}
 \hPsi^n(t,\bx+\ba)
 &= \hT_{\ba}^\dag [\rme^{-\rmi q_n \alpha} \hPsi^n,A_\mu]
 \rme^{-\rmi q_n \alpha (x)}
 \hPsi^n(t,\bx) 
 \hT_{\ba}[\rme^{-\rmi q_n \alpha} \hPsi^n,A_\mu]
 \notag \\
 &= \hT_{\ba}^\dag [\hPsi^n,A_\mu+\partial_\mu \alpha]
 \rme^{-\rmi q_n \alpha (x)}
 \hPsi^n(t,\bx) 
 \hT_{\ba}[\hPsi^n,A_\mu+\partial_\mu \alpha]
 \notag \\
 &= \rme^{-\rmi q_n \alpha (x)}
 \rme^{-\rmi q_n f_{A+\diff \alpha} (\bx_\perp+\ba_\perp,\bx_\perp)}
 \hPsi^n(t,\bx + \ba) ,
\end{align}
\endgroup
where we have explicitly shown the gauge dependence of the phase function with the subscript $A$.
Comparing both sides of this equation, we find the gauge transformation rule of the phase factor $\rme^{- \rmi q_n f (\bx_\perp+\ba_\perp,\bx_\perp)}$ as%
% \footnote{
% This transformation law can also be checked directly from Eq.~\eqref{eq:finite-f}. 
% }
% \begin{equation}
%  \U(1)_\mathrm{local}:~    
%  f (\bx_\perp+\ba_\perp,\bx_\perp)
%  \to
%  f (\bx_\perp+\ba_\perp,\bx_\perp)
%  + \alpha (t,\bx + \ba) - \alpha (t,\bx).
%  \label{eq:gauge-transformation-f}
% \end{equation}
% As a result, the gauge transformation acts on the phase factor as
\begin{equation}
 \U(1)_\mathrm{local}:~    
 \rme^{-\rmi q_n f (\bx_\perp+\ba_\perp,\bx_\perp)} 
 \to 
 \rme^{-\rmi q_n f (\bx_\perp+\ba_\perp,\bx_\perp) - \rmi q_n [\alpha (t,\bx + \ba) - \alpha (t,\bx)]} .
 \label{eq:gauge-transformation-f}
\end{equation}
This is precisely the gauge transformation law of a Wilson line for a particle with charge $q_n$ connecting $\bx$ and $\bx+\ba$.

We can also identify the gauge transformation rule of $g_{\theta} (\bx_\perp)$.
Repeating the same argument for the magnetic rotation, we find the gauge transformation rule of $\rme^{-\rmi q_n g_\theta(\bx_\perp)}$ as
\begin{equation}
 \U(1)_\mathrm{local}:~
 \rme^{-\rmi q_n g_\theta(\bx_\perp)}
 \to
 \rme^{
  - \rmi q_n g_\theta(\bx_\perp)
  - \rmi q_n [\alpha(t,R_\theta\bx) - \alpha(t,\bx)]
 }.
\end{equation}
This coincides with the gauge transformation law of a Wilson line for a particle with charge $q_n$ connecting $\bx$ and $R_\theta\bx$.

\section{Constraints on correlation functions}
\label{sec:correlation-function}

In this section, we investigate the consequences of the magnetic symmetries for correlation functions.
We first focus on two-point functions and derive the selection rules imposed by the $\U(1)$ and magnetic rotation symmetries in Sec.~\ref{sec:selection-rule}.
We then show in Sec.~\ref{sec:schwinger-phase} that magnetic translation symmetry forces two-point functions of charged operators to acquire a Schwinger phase~\cite{Schwinger:1951nm}.
In Sec.~\ref{sec:higher-point}, we finally generalize these symmetry constraints to higher-point functions.

\subsection{Selection rule}
\label{sec:selection-rule}

We begin by deriving the constraints imposed by the $\U(1)$ and magnetic rotation symmetries.

Let us consider the ground-state two-point Wightman function of local operators $\hOcal_a(x_1)$ and $\hOcal_b^\dagger(x_2)$ with $x_n=(t_n,\bx_n)$ for $n=1,2$, carrying $\U(1)$ charges $q_a$ and $q_b$ and intrinsic spins $s_a$ and $s_b$ under $\SO(2)_z$, respectively:
\begin{equation}
 G_{ab}(x_1, x_2) := 
 \bra{\Omega} 
 \hOcal_a (x_1)
 \hOcal_b^\dagger(x_2) 
 \ket{\Omega},
 \label{eq:def-two-point-function}
\end{equation}
where $\ket{\Omega}$ denotes the ground state in the presence of the
magnetic field.
In this section, we assume that the ground state preserves the time-translation,
longitudinal translation, magnetic translation, magnetic rotation, and global
$\U(1)$ symmetries and satisfies
\begin{equation}
 \hH \ket{\Omega} = 0 ,
 \quad
 \hP_z \ket{\Omega} = 0 ,
 \quad
 \hK_i \ket{\Omega} = 0,
 \quad
 \hQ \ket{\Omega} = 0,
 \quad
 \hJ_z \ket{\Omega} = j_{0}\ket{\Omega} .
 \label{eq:assumption-no-SSB}
\end{equation}
where we choose the ground-state energy to be zero.
Here, we allow a nonvanishing angular momentum $j_0$, since the background magnetic field may polarize the ground state even without spontaneous symmetry breaking.

Inserting the identity operator $1 = \hU(\alpha) \hU^\dag(\alpha)$ into Eq.~\eqref{eq:def-two-point-function}, we find the following equality:
\begin{equation}
 \begin{split}
 G_{ab}(x_1, x_2) 
%  &= 
%  \bra{\Omega}
%  U(\alpha) U^\dag(\alpha) 
%  \hOcal_a(t,\bx)
%  U(\alpha) U^\dag(\alpha)
%  \hOcal_b^\dagger(t',\by) 
%  U(\alpha) U^\dag(\alpha) \ket{\Omega}
%  \\
 &= \rme^{ \rmi (q_a - q_b) \alpha}
 G_{ab}(x_1, x_2) ,
 \end{split}
\end{equation}
where we used Eqs.~\eqref{eq:operator-phase-transformation} and \eqref{eq:assumption-no-SSB}.
This equality implies the usual selection rule for the $\U(1)$ charge as
\begin{equation}
 q_a - q_b = 0.
\end{equation}
We next insert $1 = \hR(\theta) \hR^\dag(\theta)$ to Eq.~\eqref{eq:def-two-point-function} to obtain
\begin{equation}
 \begin{split}
 G_{ab}(x_1, x_2) 
%  &= 
%  \bra{\Omega}
%  R(\theta) R^\dag(\theta)
%  \hOcal_a(t,\bx)
%  R(\theta) R^\dag(\theta)
%  \hOcal_b^\dagger(t',\by) 
%  R(\theta) R^\dag(\theta) 
%  \ket{\Omega}
%  \\
 &= \rme^{-\rmi q_a [g_\theta (\bx_{1\perp}) - g_\theta (\bx_{2\perp})]}
 \rme^{\rmi (s_a -s_b) \theta}
 G_{ab} (R_\theta x_1, R_\theta x_2),
 \end{split}
\end{equation}
where we used $q_a = q_b$ and a shorthand notation $R_\theta x = (t,R_\theta \bx)$.
Expanding this equality with respect to $\theta$, we obtain the constraint associated with magnetic rotation symmetry:
\begin{equation}
 0 = \left[
 \big(
 \bx_{1\perp} \times \bD_{x_1}
 \big)_z
 +
 \big(
 \bx_{2\perp} \times \bD_{x_2}^\dag
 \big)_z
 +
 \rmi(s_a-s_b)
 + \frac{\rmi q_a B}{2} (\bx_{1\perp}^2 - \bx_{2\perp}^2 )
 \right]
 G_{ab}(x_1,x_2) .
\end{equation}
Note that the differential operator appearing in this equation represents the total generator $\hJ_z^{(x_1)} + \hJ_z^{(x_2)}$ on the two-point function. 
Therefore, this constraint expresses the conservation of the total angular momentum associated with the magnetic rotation symmetry.

\subsection{Appearance of the Schwinger phase}
\label{sec:schwinger-phase}

We next consider the consequences of the translation symmetries.
Inserting $\hUcal \hUcal^\dag =1$ and applying the spacetime translation rule \eqref{eq:spacetime-translation} to the two-point function \eqref{eq:def-two-point-function}, we find
\begin{align}
 G_{ab}(x_1,x_2) 
 &= 
 \bra{\Omega} 
  \hUcal (-x_1) \hUcal^\dag (-x_1) 
  \hOcal_a(x_1)
  \hUcal (-x_1) \hUcal^\dag (-x_1) 
 \notag \\
 &\hspace{30pt}
  \times \hUcal (-x_2) \hUcal^\dag (-x_2) 
  \hOcal_b^\dagger(x_2) 
  \hUcal (-x_2) \hUcal^\dag (-x_2) 
 \ket{\Omega}
 \notag \\
 &= \rme^{-\rmi q_a [f(0,\bx_{1\perp}) - f (0,\bx_{2\perp})]}
 \bra{\Omega} 
  \hOcal_a(0)
  \hUcal (x_1) 
  \hUcal (-x_2) 
  \hOcal_b^\dagger(0) 
 \ket{\Omega}
 \notag \\
 % &= \rme^{
 %    -\rmi q_a [f(0,\bx_{1\perp}) - f (0,\bx_{2\perp})]
 %   }
 % \bra{\Omega} 
 %  \rme^{\frac{\rmi}{2} F_{ij} x_{1\perp}^i x_{2\perp}^j \hQ}
 %  \hOcal_a(0)
 %  \rme^{-\frac{\rmi}{2} F_{ij} x_{1\perp}^i x_{2\perp}^j \hQ}
 %  \hUcal (x_1 - x_2) 
 %  \hOcal_b^\dagger(0) 
 % \ket{\Omega}.
 % \notag \\
 % &= \rme^{
 %    -\rmi q_a [f(0,\bx_{1\perp}) - f (0,\bx_{2\perp})]
 %    - \rmi q_a \frac{1}{2} F_{ij} x_{1\perp}^i x_{2\perp}^j
 %   }
 % \bra{\Omega} 
 %  \hOcal_a(0)
 %  \hUcal (x_1 - x_2) 
 %  \hOcal_b^\dagger(0) 
 % \ket{\Omega}.
 % \notag \\
 &= \rme^{
    - \rmi q_a \Phi (\bx_{1\perp},\bx_{2\perp}) 
   }
 \bra{\Omega} 
  \hUcal^\dag (x_1 - x_2) 
  \hOcal_a(0)
  \hUcal (x_1 - x_2) 
  \hOcal_b^\dagger(0) 
 \ket{\Omega}.
 \label{eq:2-point-factorized-form}
\end{align}
Here, we used Eqs.~\eqref{eq:projective-spacetime-translation} and \eqref{eq:def-charge} together with $q_a = q_b$, and defined the phase
\begin{align}
 \Phi (\bx_{1\perp},\bx_{2\perp}) 
 := f (0,\bx_{1\perp})
  - f (0,\bx_{2\perp})
  + \frac{1}{2}F_{ij} x_{1\perp}^i x_{2\perp}^j
  % = f (\bx_{1\perp},\bx_{2\perp})
  % + \frac{1}{2} F_{ij} (-x_{1\perp}^i) 
  % (x_{1\perp}^j - x_{2\perp}^j)
  % - \frac{1}{2}F_{ij} x_{1\perp}^i x_{2\perp}^j
\label{eq:Phi-definition}
\end{align}
Equation~\eqref{eq:2-point-factorized-form} factorizes the two-point function into a part depending only on the relative coordinate $x_1-x_2$ and a phase factor $\rme^{-\rmi q_a \Phi}$ depending separately on $\bx_{1\perp}$ and $\bx_{2\perp}$.

Recalling the gauge transformation rule of $f(\bx_\perp+\ba_\perp, \bx_\perp)$ in Eq.~\eqref{eq:gauge-transformation-f}, one finds that the phase factor $\rme^{-\rmi q_a\Phi(\bx_{1\perp},\bx_{2\perp})}$ transforms as
\begin{equation}
 \U(1)_{\mathrm{local}}:~
 \rme^{ - \rmi q_a \Phi (\bx_{1\perp},\bx_{2\perp})} 
 \to 
 \rme^{ - \rmi q_a \Phi (\bx_{1\perp},\bx_{2\perp})
 + \rmi q_a [\alpha (x_1) - \alpha (x_2)]} .
\end{equation}
This is exactly the gauge transformation rule of a Wilson line of a particle with charge $q_a$ connecting $\bx_1$ and $\bx_2$.
The phase $\Phi(\bx_{1\perp},\bx_{2\perp})$ thus carries precisely the gauge dependence required for the two-point function of charged operators.

While the definition of $\Phi$ in Eq.~\eqref{eq:Phi-definition} involves a pair of the phase functions $f$, we can identify $\Phi$ as the standard form of the Schwinger phase found in the literature. 
Indeed, we rewrite it as a line integral of $A_i$ and $F_{ij}$ by substituting the integral representation~\eqref{eq:finite-f} of $f$ into Eq.~\eqref{eq:Phi-definition} and using the composition rule~\eqref{eq:composition-rule-f} to obtain
\begin{equation}
 \Phi (\bx_{1\perp},\bx_{2\perp})
 = -\int_{\bx_{2\perp}}^{\bx_{1\perp}}
   \diff \xi^i
   \left[
    A_i(\bxi) + \frac{1}{2} F_{ij} (\xi^j - x_{2\perp}^j)
   \right] .
 \label{eq:Schwinger-line-integral}
\end{equation}
Note that the integrand here is curl-free, so that the line integral is independent of the path connecting $\bx_{2\perp}$ to $\bx_{1\perp}$.

Equation~\eqref{eq:Schwinger-line-integral} is precisely the well-known Schwinger phase~\cite{Schwinger:1951nm} appearing in the propagator of charged particles in a background magnetic field.
It is worth emphasizing that the Schwinger phase emerges solely from the projective structure of the magnetic translation symmetry without relying on a specific model or perturbative approximation.
Thus, the factorized form in Eqs.~\eqref{eq:2-point-factorized-form}-\eqref{eq:Phi-definition} holds non-perturbatively whenever the magnetic translation symmetry is unbroken.
Moreover, the representation \eqref{eq:Schwinger-line-integral} also makes the Wilson-line interpretation transparent: the line integral of $A_i$ provides the gauge-dependent Wilson-line factor, while the $F$-dependent correction is manifestly gauge invariant.

\subsection{Generalization to higher-point functions}
\label{sec:higher-point}

We finally extend the analysis to derive the selection rules and the Schwinger-phase factorization for the higher-point functions.

Let us consider the $N$-point function of local operators,
\begin{equation}
 G_{a_1\cdots a_N}(x_1,\dots,x_N)
 :=
 \bra{\Omega}
 \hOcal_{a_1}(x_1)\,
 \hOcal_{a_2}(x_2)\cdots
 \hOcal_{a_N}(x_N)
 \ket{\Omega},
 \label{eq:def-N-point-function}
\end{equation}
where $\hOcal_{a_n}$ carries $\U(1)$ charge $q_{a_n}$ and intrinsic spin $s_{a_n}$ under $\SO(2)_z$.

\paragraph{Selection rules}

We first generalize the selection rules.
Inserting $\hU(\alpha) \hU^\dagger(\alpha)=\bm{1}$ into the $N$-point function and using Eq.~\eqref{eq:operator-phase-transformation}, we find
\begin{equation}
 G_{a_1\cdots a_N}(x_1,\dots,x_N)
 =
 \rme^{\rmi \alpha \sum_{n=1}^{N} q_{a_n}}
 G_{a_1\cdots a_N}(x_1,\dots,x_N).
\end{equation}
Thus, the $N$-point function can be nonzero only if
\begin{equation}
 \sum_{n=1}^{N} q_{a_n}=0.
 \label{eq:N-point-charge-selection}
\end{equation}

Similarly, inserting $\hR(\theta) \hR^\dagger(\theta)=\bm{1}$ and using Eq.~\eqref{eq:operator-magnetic-rotation}, we obtain the magnetic-rotation constraint.
Expanding to first order in $\theta$, we find
\begin{equation}
 \sum_{n=1}^{N}
 \left[
 \big(\bx_{n\perp}\times\bD_{x_n}\big)_z
 + \rmi s_{a_n}
 + \frac{\rmi q_{a_n} B}{2}\,\bx_{n\perp}^2
 \right]
 G_{a_1\cdots a_N}(x_1,\dots,x_N)
 =
 0.
 \label{eq:N-point-rotation-selection}
\end{equation}
Here, $\bD_{x_n}$ is understood to act on the operator with charge $q_{a_n}$.
Equation~\eqref{eq:N-point-rotation-selection} expresses the conservation of the total magnetic angular momentum acting on the $N$-point function.

\paragraph{Schwinger-phase factorization}

We next consider the consequence of magnetic translation symmetry.
As in the two-point case, we translate each operator to the origin by inserting
\[
 \hUcal(-x_n)\hUcal^\dagger(-x_n)=\bm{1}.
\]
Using the spacetime translation rule~\eqref{eq:spacetime-translation}, we have
\begin{equation}
 \hUcal^\dagger(-x_n)\,
 \hOcal_{a_n}(x_n)\,
 \hUcal(-x_n)
 =
 \rme^{-\rmi q_{a_n} f(0,\bx_{n\perp})}
 \hOcal_{a_n}(0).
\end{equation}
The neighboring translation operators combine according to the projective composition law~\eqref{eq:projective-spacetime-translation} as
\begin{equation}
 \hUcal(x_n)\hUcal(-x_{n+1})
 =
 \rme^{-\frac{\rmi}{2}F_{ij}x_{n\perp}^i x_{n+1\perp}^j \hQ}
 \hUcal(x_n-x_{n+1}).
\end{equation}
Moving the central phase factors to the right through the operators at the origin produces charge-dependent c-number phases.
As a result, the $N$-point function factorizes as
\begin{equation}
 G_{a_1\cdots a_N}(x_1,\dots,x_N)
 =
 \rme^{-\rmi\Theta_N}\,
 \bar G_{a_1\cdots a_N} (x_1-x_2,\dots,x_{N-1}-x_N),
 \label{eq:N-point-factorized}
\end{equation}
where we defined the reduced correlator by
\begin{align}
 \bar G_{a_1\cdots a_N} (x_1-x_2,\dots,x_{N-1}-x_N)
 &:=
 \bra{\Omega}
 \hOcal_{a_1}(0)\,
 \hUcal(x_1-x_2)\,
 \hOcal_{a_2}(0)\,
 \hUcal(x_2-x_3)\cdots
 \notag\\
 &\hspace{35pt}\times
 \hUcal(x_{N-1}-x_N)\,
 \hOcal_{a_N}(0)
 \ket{\Omega}.
 \label{eq:N-point-reduced}
\end{align}
According to the above procedure, the phase $\Theta_N$ appearing in Eq.~\eqref{eq:N-point-factorized} is decomposed into two parts:
\begin{equation}
 \Theta_N
 =
 \Theta_{\mathrm{gauge}}
 +
 \Theta_{\mathrm{inv}},
 \label{eq:N-point-phase}
\end{equation}
with
\begin{equation}
 \Theta_{\mathrm{gauge}}
 =
 \sum_{n=1}^{N}
 q_{a_n}\, f(0,\bx_{n\perp}),
 \qquad
 \Theta_{\mathrm{inv}}
 =
 -\frac{1}{2}
 \sum_{n=1}^{N-1}
 F_{ij}\,x_{n\perp}^i x_{n+1\perp}^j\,
 \mathcal{Q}_n.
 \label{eq:N-point-phase-parts}
\end{equation}
Here, we defined the total $\U(1)$ charge carried by the operators to the right of the $n$-th link factor:
\begin{equation}
 \mathcal{Q}_n
 :=
 \sum_{m=n+1}^{N} q_{a_m}.
\end{equation}
Thus, magnetic translation symmetry fixes the coordinate dependence of the $N$-point function up to the reduced correlator~\eqref{eq:N-point-reduced}, which depends only on coordinate differences.

Let us comment on gauge covariance.
Under a local $\U(1)$ transformation, the phase function transforms as in Eq.~\eqref{eq:gauge-transformation-f}, and the charge selection rule~\eqref{eq:N-point-charge-selection} ensures that
\begin{equation}
 \U(1)_{\mathrm{local}}:~
 \Theta_{\mathrm{gauge}}
 \to
 \Theta_{\mathrm{gauge}}
 -
 \sum_{n=1}^{N} q_{a_n}\alpha(x_n).
\end{equation}
Therefore, the factor $\rme^{-\rmi\Theta_N}$ supplies precisely the gauge dependence required for the $N$-point function of charged operators.

The two terms in $\Theta_N$ have a simple geometric interpretation.
The gauge-dependent part $\Theta_{\mathrm{gauge}}$ is a sum of Wilson-line phases connecting each insertion point to the reference point.
The gauge-invariant part is built only from the field strength.
Indeed, defining the oriented magnetic flux through the triangle with vertices $\bzero$, $\bx_{n\perp}$, and $\bx_{n+1\perp}$ by
\begin{equation}
 \Phi_{\triangle B} (\bzero,\bx_{n\perp},\bx_{n+1\perp})
 :=
 \frac{1}{2}
 F_{ij}\,x_{n\perp}^i x_{n+1\perp}^j,
\end{equation}
we can express
\begin{equation}
 \Theta_{\mathrm{inv}}
 =
 -\sum_{n=1}^{N-1}
 \mathcal{Q}_n\,
 \Phi_{\triangle B} (\bzero,\bx_{n\perp},\bx_{n+1\perp}).
 \label{eq:N-point-flux}
\end{equation}
Thus, the gauge-invariant part of the $N$-point Schwinger phase is a sum of magnetic fluxes through oriented triangles, weighted by the partial charges $\mathcal{Q}_n$.
The decomposition into triangular fluxes in Eq.~\eqref{eq:N-point-flux} parallels the polygon Schwinger phase identified in the diagrammatic analysis of multi-point correlators in a uniform magnetic field~(see, e.g., Ref.~\cite{Hattori:2023egw}).
Our derivation shows that this structure follows directly from magnetic translation symmetry, without relying on a perturbative expansion.

For $N=2$, we have $\mathcal{Q}_1=q_{a_2}=-q_a$ after imposing the charge selection rule.
Then Eqs.~\eqref{eq:N-point-factorized}--\eqref{eq:N-point-phase-parts} reduce to the two-point factorization formula~\eqref{eq:2-point-factorized-form}, with $\Theta_2=q_a\Phi(\bx_{1\perp},\bx_{2\perp})$.

\section{Spectral representation in magnetic fields}
\label{sec:spectral}

In the previous section, we have found that the two-point Wightman function takes the factorized form \eqref{eq:2-point-factorized-form} as the Schwinger phase and a reduced correlator 
\begin{equation}
 G_{ab}(x_1,x_2)
 =
 \rme^{-\rmi q_a \Phi(\bx_{1\perp},\bx_{2\perp})}\,
 \bar G_{ab}(x_1-x_2),
\end{equation}
where we defined the reduced correlator by
\begin{equation}
 \bar G_{ab}(x)
 :=
 \bra{\Omega}
 \hUcal^\dag (x)
 \hOcal_a(0)
 \hUcal(x)
 \hOcal_b^\dagger(0)
 \ket{\Omega}.
 \label{eq:reduced-correlator}
\end{equation}
The reduced correlator is gauge invariant and depends only on the relative coordinate.

In this section, we derive the spectral representation of the reduced two-point function implied by magnetic translation symmetry.
After formulating the general spectral decomposition of the reduced correlator in terms of representations of the magnetic translation algebra in Sec.~\ref{sec:reduced-correlator-magnetic-rep}, we construct explicit projected spectral representations adapted to several standard bases in Sec.~\ref{sec:projected-spectral-representations}.
We finally discuss how other Green's functions are reconstructed from the same spectral data in Sec.~\ref{sec:other-green-functions}.

\subsection{Representations of the magnetic translation symmetry}
\label{sec:reduced-correlator-magnetic-rep}

In ordinary translationally invariant systems, spectral representations are organized in terms of simultaneous eigenstates of mutually commuting momentum operators.
In the present case, the longitudinal momentum $P_z$ remains a good quantum number, whereas the transverse magnetic translation generators obey the centrally extended algebra~\eqref{eq:KK-commutator}.
Therefore, in a charge-$q$ sector, the magnetic translation algebra reduces to
 \begin{equation}
 [\hK_i,\hK_j]\big|_q
 = -\rmi F_{ij} q .
 \label{eq:magnetic-translation-heisenberg}
\end{equation}
For neutral sectors with $q=0$, this algebra becomes Abelian, and the magnetic translation generators reduce to ordinary commuting transverse momentum operators.
In this case, the spectral representation takes the ordinary momentum-space form, with plane-wave eigenstates of $\hK_1, \hK_2$ labeled by transverse momenta $\bp_\perp$.

In contrast, the magnetic translation generators for $q\neq0$ become noncommutative, which forbids us to diagonalize $\hK_1$ and $\hK_2$ simultaneously.
Although the reduced correlator $\bar G_{ab}(x)$ can be Fourier transformed as a function of the relative coordinate, its symmetry-adapted spectral representation is not organized by transverse momentum eigenstates.
Instead, the transverse dependence is encoded in representation matrices of the magnetic translation algebra.

For each charge sector with $q \neq 0$, one convenient way to realize the noncommutative algebra is to introduce
\begin{equation}
 \ha_q
 :=
 \frac{\hK_1-\rmi \operatorname{sgn}(qB) \hK_2}{\sqrt{2|qB|}},
 \qquad
 \ha_q^\dagger
 :=
 \frac{\hK_1+\rmi \operatorname{sgn}(qB) \hK_2}{\sqrt{2|qB|}} ,
 \label{eq:def-magnetic-oscillator}
\end{equation}
which satisfy
\begin{equation}
 \big[\ha_q, \ha_q^\dagger \big] = 1.
\end{equation}
For each fixed set of quantum numbers commuting with the magnetic translation generators, the representation space of magnetic translations can be realized by a harmonic-oscillator-type basis.
We emphasize that this oscillator structure follows solely from the magnetic translation algebra and should not be identified with the cyclotron dynamics of a particular microscopic model.

We now insert a complete set of states,
\begin{equation}
 \bm{1}
 = \sum_q \sum_{\lambda,\alpha} \int \frac{\diff p_z}{2\pi}\,
 \ket{\lambda,p_z,\alpha;q}
 \bra{\lambda,p_z,\alpha;q}. \label{eq:complete-set-charge-q}
\end{equation}
Here, $\lambda$ collectively denotes quantum numbers other than the longitudinal momentum and the magnetic-translation degeneracy index, including any additional labels commuting with $\hK_i$.
The label $\alpha$ specifies a basis of the magnetic-translation representation space and may be either discrete or continuous, depending on the chosen basis.
The summation symbols should be understood to include integrations over continuous labels whenever necessary.
We choose these states to satisfy
\begin{subequations} \label{eq:state-quantum-numbers}
\begin{align}
 \hH \ket{\lambda,p_z,\alpha;q}
 &=
 E_\lambda^{(q)} (p_z) \ket{\lambda,p_z,\alpha;q},
 \\
 \hP_z \ket{\lambda,p_z,\alpha;q}
 &=
 p_z \ket{\lambda,p_z,\alpha;q},
 \\
 \hQ \ket{\lambda,p_z,\alpha;q}
 &=
 q\ket{\lambda,p_z,\alpha;q}.
\end{align}
\end{subequations}
Here, the index $\alpha$ labels states within the same $(\lambda, p_z)$ sector that are related by magnetic translations.
Since magnetic translations commute with $\hH$, the energy $E^{(q)}_\lambda(p_z)$ is independent of $\alpha$.

The operator $\hT(\bx_\perp)$ then acts within each fixed $\lambda, p_z, q$ sector as
\begin{equation}
 \hT(\bx_\perp)
 \ket{\lambda,p_z,\alpha;q}
 = \sum_\beta
 D_{\beta\alpha}^{(q)}(\bx_\perp)
 \ket{\lambda,p_z,\beta;q},
 \label{eq:D-matrix-definition}
\end{equation}
where we introduced the representation matrix by
\begin{equation}
 D_{\beta\alpha}^{(q)}(\bx_\perp) 
 \, \delta_{\lambda'\lambda}
 \, 2\pi \delta (p_z' - p_z)
 :=
 \bra{\lambda',p_z',\beta;q}
 \hT(\bx_\perp)
 \ket{\lambda,p_z,\alpha;q}.
 \label{eq:D-matrix-element}
\end{equation}
For notational simplicity, we suppress the $(\lambda,p_z)$ labels on the representation matrices $D^{(q)}$, since the magnetic translation operators act only on the indices $\alpha,\beta$.
The matrices $D^{(q)}(\bx_\perp)$ furnish a representation of the magnetic translation group on the magnetic-translation representation space.
They play the role analogous to the plane-wave factor $\rme^{\rmi \bp\cdot\bx}$ appearing in ordinary translations as $\hT(\bx)\ket{\bp}|_{B=0} = \rme^{\rmi \bp\cdot\bx} \ket{\bp}$.

The representation matrices $D^{(q)}$ satisfy the following four conditions. 
First, we have the unitarity condition and the projective composition law following from Eq.~\eqref{eq:projective-rep.}:
\begin{align}
 &\text{(i)~Unitarity~condition:}~
 D_{\beta\alpha}^{(q)}(\bx_\perp)^*
 = D_{\alpha\beta}^{(q)}(-\bx_\perp)
 ,
 \label{eq:D-unitary}
 \\
 &\text{(ii)~Projective~composition:}~
 \sum_{\gamma}
 D_{\alpha\gamma}^{(q)}(\bx_\perp) D_{\gamma\beta}^{(q)}(\by_\perp)
 = \rme^{ \frac{\rmi}{2} q F_{ij} x_\perp^i y_\perp^j}
 D_{\alpha\beta}^{(q)}(\bx_\perp+\by_\perp).
 \label{eq:D-projective}
\end{align}
Second, by choosing suitably normalized irreducible bases, one may impose the orthogonality relation
\begin{equation}
 \text{(iii)~Orthogonality~condition:}~
 \int \diff^2 x_\perp \,
 D^{(q)*}_{\alpha\beta}(\bx_\perp) D^{(q)}_{\alpha'\beta'}(\bx_\perp)
 = \frac{2\pi}{|qB|}
 \delta_{\alpha\alpha'} \delta_{\beta\beta'},
 \label{eq:D-orthogonality} 
\end{equation}
where the Kronecker deltas in the orthogonality condition should be understood as Dirac deltas when the representation labels $\alpha,\beta$ are taken to be continuous ones.

Moreover, the magnetic translation representations in the charge $\pm q$ sectors are related by complex conjugation.
Indeed, taking the complex conjugate of the projective composition law in Eq.~\eqref{eq:D-projective} yields
\begin{equation}
 \sum_\gamma D^{(q)*}_{\alpha\gamma}(\bx_\perp)
 D^{(q)*}_{\gamma\beta}(\by_\perp)
 =
 \rme^{-\frac{\rmi}{2} q F_{ij} x_\perp^i y_\perp^j}
 D^{(q)*}_{\alpha\beta}(\bx_\perp+\by_\perp),
\end{equation}
which is precisely the projective composition law for the charge-$(-q)$ sector.
Choosing the basis in the charge-$(-q)$ sector as the complex-conjugate basis of the charge-$q$ sector, we obtain
\begin{equation}
 \text{(iv)~Complex~conjugation~relation:}~
 D^{(q)*}_{\alpha\beta}(\bx_\perp)
 =
 D^{(-q)}_{\alpha\beta}(\bx_\perp).
 % =
 % D^{(q)}_{\beta\alpha}(-\bx_\perp),
 \label{eq:D-complex-conjugation}
\end{equation}

Thus, up to a choice of basis for the representation index $\alpha$, the transverse dependence carried by $D^{(q)}$ is constrained by the magnetic translation algebra.

\subsection{Projected spectral representations}
\label{sec:projected-spectral-representations}

We then insert the complete set~\eqref{eq:complete-set-charge-q} twice into Eq.~\eqref{eq:reduced-correlator} and simplify the result.
The matrix element is nonvanishing only when $q_a=q_b$, as required by
the $\U(1)$ selection rule. 
We then obtain
\begin{align}
 \bar{G}_{ab}(x)
%  &= \sum_{\lambda,\lambda',\alpha,\beta}
% \int \frac{\diff p_z}{2\pi}
% \int \frac{\diff p_z'}{2\pi} 
% \bra{\Omega}
%  \hOcal_a(0)
%  \ket{\lambda,p_z,\alpha;q}
%  \bra{\lambda,p_z,\alpha;q}
%  \hUcal(x)
%  \ket{\lambda',p_z',\beta;q}
%  \bra{\lambda',p_z',\beta;q}
%  \hOcal_b^\dagger(0)
%  \ket{\Omega}
%  \notag \\
 &=
\sum_{\lambda,\alpha,\beta} 
\int \frac{\diff p_z}{2\pi}
 \rme^{-\rmi E^{(q_a)}_\lambda(p_z) t + \rmi p_z z}
 \bra{\Omega}
 \hOcal_a(0)
 \ket{\lambda,p_z,\alpha;q_a}
 \hspace{-1pt}
 D_{\alpha\beta}^{(q_a)} (\bx_{\perp})
 \hspace{-1pt}
 \bra{\lambda,p_z,\beta;q_a}
 \hOcal_b^\dagger(0)
 \ket{\Omega},
\end{align}
where we used the decomposition of the spacetime translation operator in Eq.~\eqref{eq:translation-op-decomposition}, together with Eq.~\eqref{eq:D-matrix-element}.
As a result, we arrive at
\begin{equation}
 \bar G_{ab}(x)
 =
 \sum_\lambda
 \int \frac{\diff p_z}{2\pi}
 \rme^{-\rmi E^{(q_a)}_\lambda (p_z) t+\rmi p_z z}
 \sum_{\alpha,\beta}
 \Acal^{(a)}_\alpha(\lambda,p_z)\,
 D_{\alpha\beta}^{(q_a)}(\bx_\perp)\,
 \Acal^{(b)*}_\beta(\lambda,p_z),
 \label{eq:magnetic-spectral}
\end{equation}
with the matrix elements 
\begin{equation}
 \mathcal A^{(c)}_\alpha(\lambda,p_z)
 := \bra{\Omega} \hOcal_c (0) \ket{\lambda,p_z,\alpha;q_c}
 % \quad \mathrm{and} \quad
 % \mathcal A^{(b)*}_\beta(\lambda,p_z)
 % := \bra{\lambda,p_z,\beta;q_b} \hOcal_b^\dag (0) \ket{\Omega}.
 \label{eq:spectral-amplitude}
\end{equation}
and its complex conjugate for $c=a,b$.

Equations~\eqref{eq:magnetic-spectral}-\eqref{eq:spectral-amplitude} give the spectral representation dictated by the magnetic translation symmetry.
Note that the transverse coordinate dependence is carried by the representation matrix $D_{\alpha\beta}^{(q)}(\bx_\perp)$, which replaces the plane wave
$\rme^{\rmi \bp_\perp\cdot\bx_\perp}$ appearing in ordinary translationally invariant systems.
The dynamical information is encoded in the excitation spectrum $\{E^{(q)}_\lambda(p_z)\}$ and the amplitudes $\mathcal A^{(c)}_\alpha(\lambda,p_z)$, while the form of $D^{(q)}$ is fixed by the magnetic translation algebra.
In this sense, the matrices $D^{(q)}(\bx_\perp)$ play the role of symmetry-adapted basis functions in the presence of a background magnetic field.

The choice of basis for the index $\alpha$ corresponds to choosing a maximal commuting set within the magnetic symmetry algebra.
For example, diagonalizing one magnetic translation generator leads to the basis naturally associated with the Landau-gauge description, whereas diagonalizing the magnetic rotation generator $\hJ_z$ leads to the rotational basis familiar from the symmetric-gauge description.
These correspond to different basis choices of the same irreducible representation of the magnetic translation algebra and do not affect the basis-independent form of Eq.~\eqref{eq:magnetic-spectral}.

To obtain a more explicit spectral representation analogous to the ordinary momentum-space representation, it is convenient to project the reduced correlator onto a chosen basis of the magnetic-translation representation space.
For a given basis labeled by $\alpha, \beta$, we define the projected correlator by
\begin{equation}
 \bar G^{(ab)}_{\alpha\beta}(E, p_z)
 :=
 \frac{|q_a B|}{2\pi}
 \int \diff t \, \diff z \, \diff^2 x_\perp \,
 \rme^{\rmi E t - \rmi p_z z}
 D_{\alpha\beta}^{(q_a)*}(\bx_\perp)
 \bar G_{ab}(x).
 \label{eq:projected-green-general}
\end{equation}
Substituting Eq.~\eqref{eq:magnetic-spectral} into this definition and using the orthogonality relation~\eqref{eq:D-orthogonality}, we obtain
\begin{equation}
 \bar G^{(ab)}_{\alpha\beta}(E, p_z)
 =
 2\pi \sum_\lambda
 \delta(E - E^{(q_a)}_\lambda(p_z))
 \mathcal A^{(a)}_\alpha(\lambda, p_z)
 \mathcal A^{(b)*}_\beta(\lambda, p_z).
 \label{eq:projected-green-spectral}
\end{equation}
This is the magnetic-field analogue of the ordinary spectral representation in momentum space~\cite{Umezawa:1951rp,Kallen:1952zz,Lehmann:1954xi}, with the transverse momentum eigenstates replaced by the magnetic-translation representation index $\alpha$.

The required form depends on the choice of basis for the representation index.
Below we describe two standard choices.

\paragraph{(a) $K_y$-diagonal basis}

One convenient choice is to diagonalize one magnetic translation generator, e.g. $\hK_y$.
We denote its eigenvalue by $k_y$,
\begin{equation}
 \hK_y \ket{\lambda,p_z,k_y;q}
 =
 k_y \ket{\lambda,p_z,k_y;q}.
\end{equation}
Since $[\hK_x,\hK_y] \big|_q = - \rmi B q$, the remaining generator $\hK_x$ acts as the conjugate operator, $\hK_x = - \rmi q B \frac{\partial}{\partial k_y}$.
Thus, the representation is naturally realized on functions of the continuous variable $k_y$.
This basis is naturally associated with the Landau-gauge description, in which one transverse translation generator remains manifestly diagonal.

In the $K_y$-diagonal basis, the magnetic translation matrix is defined by
\begin{equation}
 \hT(\bx_\perp)
 \ket{\lambda,p_z,k_y';q}
 =
 \int \diff k_y
 D_{k_y k_y'}^{(q)}(\bx_\perp)
 \ket{\lambda,p_z,k_y;q}.
\end{equation}
Recalling the magnetic translation operator in Eq.~\eqref{eq:magnetic-translation-operator}, we obtain
\begin{align}
 \hT(\bx_\perp)
 \ket{\lambda,p_z,k_y';q}
 &=
 \rme^{-\frac{\rmi}{2}qBxy}
 \rme^{\rmi yk_y'}
 \ket{\lambda,p_z,k_y'-qBx;q}.
\end{align}
Therefore,
\begin{equation}
 D_{k_y k_y'}^{(q)}(x,y)
 = \rme^{-\frac{\rmi}{2}qBxy}
 \rme^{\rmi yk_y'}
 \delta(k_y-k_y'+qBx)
 = \rme^{ \frac{\rmi}{2}y (k_y+k_y')}
 \delta(k_y-k_y'+qBx),
 \label{eq:D-Landau-basis}
\end{equation}
where the second equality follows after rewriting the phase factor using the delta function.

As expected, the phase factor takes a plane-wave form in the $y$ direction, while the delta function shifts the magnetic translation quantum number $k_y$ along the $x$ direction.
This mixed structure is characteristic of the basis commonly used in the Landau-gauge description, where one transverse translation generator is diagonalized.

The orthogonality relation \eqref{eq:D-orthogonality} in this case takes the following explicit form
\begin{equation}
 \int \diff^2x_\perp\,
 D_{k_y k_y'}^{(q)*}(\bx_\perp)
 D_{\ell_y \ell_y'}^{(q)}(\bx_\perp)
 =
 \frac{2\pi}{|qB|}
 \delta(k_y-\ell_y)
 \delta(k_y'-\ell_y'),
 \label{eq:D-Landau-orthogonality}
\end{equation}
in which the Kronecker deltas are replaced by Dirac deltas for the continuous label $k_y$.
Thus, the projected correlator \eqref{eq:projected-green-spectral} in this basis reads
\begin{equation}
 \bar G^{(ab)}_{k_y k_y'}(E,p_z)
 =
 2\pi
 \sum_\lambda
 \delta\!\left(E-E^{(q_a)}_\lambda(p_z)\right)
 \mathcal A^{(a)}_{k_y}(\lambda,p_z)
 \mathcal A^{(b)*}_{k_y'}(\lambda,p_z).
 \label{eq:projected-wightman-Landau}
\end{equation}
This is the magnetic-field analogue of the ordinary spectral representation in momentum space, with the transverse momentum replaced by the magnetic-translation quantum number $k_y$.

\paragraph{(b) $J_z$-diagonal basis}

Another useful choice is to diagonalize the magnetic rotation generator
$\hJ_z$.
Using the oscillator realization introduced in Eq.~\eqref{eq:def-magnetic-oscillator} and Eq.~\eqref{eq:JK-commutator}, we find  
\begin{equation}
 [\hJ_z,\ha_q]
 = -\operatorname{sgn}(qB) \ha_q,
 \qquad
 [\hJ_z,\ha_q^\dagger]
 = \operatorname{sgn}(qB) \ha_q^\dagger.
 \label{eq:ladder-Jz}
\end{equation}
Thus, the operators $\ha_q$ and $\ha_q^\dag$ indeed act as lowering and raising operators
for the magnetic angular momentum.
Let $\ket{\lambda,p_z,0;q}$ be a reference state satisfying
\begin{equation}
 \hJ_z \ket{\lambda,p_z,0;q}
 = j_{\mathrm{ref}} \ket{\lambda,p_z,0;q}
 \quad \mathrm{and} \quad
 \ha_q \ket{\lambda,p_z,0;q}=0.
 \label{eq:rotation-basis}
\end{equation}
We then construct the oscillator basis
\begin{equation}
 \ket{\lambda,p_z,m;q}
 :=
 \frac{(\ha_q^\dagger)^m}{\sqrt{m!}}
 \ket{\lambda,p_z,0;q}
 \with
 m=0,1,2,\cdots .
\end{equation}
Using Eq.~\eqref{eq:ladder-Jz}, one finds
\begin{equation}
 \hJ_z \ket{\lambda,p_z,m;q}
 = \left[ j_{\mathrm{ref}} + \operatorname{sgn}(qB)m \right]
 \ket{\lambda,p_z,m;q},
\end{equation}
so the oscillator number basis is simultaneously a
$J_z$-diagonal basis.
In this basis, the representation index $\alpha$ is specialized to the
oscillator number $m$.
This reproduces the rotational structure characteristic of the
symmetric-gauge description.

Restricting to a fixed charge-$q$ sector, the magnetic translation
operator can be expressed in terms of the oscillator operators as
\begin{equation}
 \hT(\bx_\perp)
 = \rme^{\xi \ha_q^\dagger-\xi^\ast \ha_q}
 = \rme^{-|\xi|^2/2}
 \rme^{\xi \ha_q^\dagger}
 \rme^{-\xi^\ast \ha_q}
 \with
 \xi
 := \rmi \sqrt{\frac{|qB|}{2}}
 \left[
  x-\rmi \operatorname{sgn}(qB)y
 \right].
\end{equation}
This operator coincides with the displacement operator of the harmonic
oscillator.
Using standard harmonic-oscillator algebra, the representation matrix \eqref{eq:D-matrix-element} in the rotational basis is evaluated as
\begin{equation}
 D_{mm'}^{(q)}(\bx_\perp)
 = \rme^{-|\xi|^2/2}
 \times
 \begin{cases}
 \ds{\sqrt{\frac{m'!}{m!}} \xi^{m-m'} L_{m'}^{(m-m')}(|\xi|^2)} ,
 & m' \leq m,
 \vspace{5pt} \\
 \ds{\sqrt{\frac{m!}{m'!}} (-\xi^\ast)^{\,m'-m} L_m^{(m'-m)}(|\xi|^2)} ,
 & m \leq m',
 \end{cases}
 \label{eq:D-rotational}
\end{equation}
where $L_m^{(\alpha)}(x)$ denotes the associated Laguerre polynomial and $m,m' =0,1,2,\cdots$.

This explicit form allows us to verify the orthogonality relation
\begin{equation}
 \int \diff^2x_\perp\,
 D_{mm'}^{(q)\,*}(\bx_\perp)
 D_{nn'}^{(q)}(\bx_\perp)
 =
 \frac{2\pi}{|qB|}
 \delta_{mn}
 \delta_{m'n'}.
 \label{eq:D-rotational-orthogonality}
\end{equation}
Thus, projecting the two-point function onto the rotational basis, we find
\begin{equation}
 \bar G^{(ab)}_{mm'}(E,p_z)
 = 2\pi
 \sum_\lambda
 \delta\!\left(E-E^{(q_a)}_\lambda(p_z)\right)
 \mathcal A^{(a)}_{m}(\lambda,p_z)
 \mathcal A^{(b)\,*}_{m'}(\lambda,p_z).
 \label{eq:projected-wightman-rotational}
\end{equation}
This gives the rotationally adapted form of the magnetic spectral representation, in which the transverse structure is organized by the magnetic angular momentum.

In summary, we find the spectral representation of the reduced Wightman correlator in Eq.~\eqref{eq:magnetic-spectral}, where the transverse coordinate dependence is encoded in the representation matrix $D^{(q)}_{\alpha\beta}(\bx_\perp)$ of the magnetic translation algebra rather than in ordinary plane waves.
Projecting onto a chosen basis of the representation space then leads to explicit spectral representations such as Eq.~\eqref{eq:projected-wightman-Landau} and Eq.~\eqref{eq:projected-wightman-rotational}.
Although their explicit forms depend on the choice of basis, they describe the same underlying excitation spectrum $E^{(q)}_\lambda(p_z)$ and the same representation of the magnetic translation algebra.

\subsection{Retarded and time-ordered Green's functions}
\label{sec:other-green-functions}

The spectral representation derived in the previous subsection applies to the Wightman correlator $\bar G_{ab}(x)$.
Other two-point Green's functions are reconstructed from the same spectral data through their standard relations to the Wightman correlator.
In this subsection, we illustrate this for the retarded and time-ordered Green's functions.

Let us first consider the retarded reduced correlator defined by
\begin{equation}
 \bar G^R_{ab}(x)
 :=
 -\rmi \theta(t)
 \bra{\Omega}
 [\hUcal^\dag(x) \hOcal_a(0) \hUcal(x), \hOcal_b^\dag(0)]_\mp
 \ket{\Omega},
 \label{eq:retarded-definition}
\end{equation}
where the upper (lower) sign corresponds to the commutator (anticommutator) for bosonic (fermionic) operators $\hOcal_a$ and $\hOcal_b^\dag$.
Using $\hUcal(x) \ket{\Omega} = \ket{\Omega}$, the retarded correlator is expressed in terms of two Wightman-type correlators:
\begin{align}
 \bar G^R_{ab}(x)
 &=
 -\rmi \theta(t)
 \left[
 \bra{\Omega} \hOcal_a(0) \hUcal(x) \hOcal_b^\dag(0) \ket{\Omega}
 \mp
 \bra{\Omega} \hOcal_b^\dag(0) \hUcal^\dag(x) \hOcal_a(0) \ket{\Omega}
 \right] .
 \label{eq:retarded-decomposition}
\end{align}
The two terms involve intermediate states in different charge sectors.
In the first term, the states are created by $\hOcal_b^\dag(0)$ acting on the ground state from the right, carrying charge $q_b = q_a$.
In the second term, they are created by $\hOcal_b^\dag(0)$ acting on the ground state from the left, carrying charge $-q_b = -q_a$.

To organize the spectral decomposition of both terms, we introduce the amplitudes and spectra for both charge sectors.
For the charge-$q_a$ sector, the energy eigenvalue $E^{(q_a)}_\lambda(p_z)$ and the amplitudes $\Acal^{(c)}_\alpha(\lambda, p_z)$ are defined in Eqs.~\eqref{eq:state-quantum-numbers} and ~\eqref{eq:spectral-amplitude}.
For the charge-$(-q_a)$ sector, we have the energy eigenvalue $E^{(-q_a)}_\lambda(p_z)$ and introduce the corresponding amplitude by
\begin{align}
 \tilAcal^{(c)}_\alpha(\lambda,p_z)
 &:= \bra{\lambda,p_z,\alpha;-q_c} \hOcal_c(0) \ket{\Omega}
 % \\
 % \tilde{\mathcal A}^{(c)*}_\alpha(\lambda,p_z)
 % &:= \bra{\Omega} \hOcal_c^\dag(0) \ket{\lambda,p_z,\alpha;-q_c}.
 \label{eq:tilde-amplitude}
\end{align}
for $c = a, b$.
We note that the spectrum $E^{(-q_a)}_\lambda(p_z)$ in the charge-$(-q_a)$ sector is generally distinct from $E^{(q_a)}_\lambda(p_z)$ in the charge-$q_a$ sector.

Inserting complete sets of states into Eq.~\eqref{eq:retarded-decomposition} twice and using the quantities defined above, we can express the projected retarded correlator as
\begin{align}
 \bar G^{(ab),R}_{\alpha\beta}(E,p_z)
 &:= \frac{|q_a B|}{2\pi}
 \int \diff t \, \diff z \, \diff^2 x_\perp \,
 \rme^{\rmi E t - \rmi p_z z}
 D^{(q_a)*}_{\alpha\beta}(\bx_\perp)
 \bar G^R_{ab}(x)
 \notag \\
 &= \frac{|q_a B|}{2\pi}
 \int \diff t \, \diff z \, \diff^2 x_\perp \,
 \rme^{\rmi E t - \rmi p_z z}
 D^{(q_a)*}_{\alpha\beta}(\bx_\perp)
 \int \frac{\diff \omega}{2\pi} \frac{-1}{\omega - \rmi 0} \rme^{\rmi \omega t}
 \notag \\
 &\qquad 
 \sum_\lambda \int \frac{\diff p_z'}{2\pi} \sum_{\alpha',\beta'}
 \bigg[
  \rme^{-\rmi E^{(q_a)}_\lambda(p_z') t + \rmi p_z' z}
  \Acal^{(a)}_{\alpha'}(\lambda,p_z') D^{(q_a)}_{\alpha'\beta'}(\bx_\perp) \Acal^{(b)*}_{\beta'}(\lambda,p_z')
 \notag \\
 &\hspace{90pt}
 \mp
  \rme^{+\rmi E^{(-q_a)}_\lambda(p_z') t - \rmi p_z' z}
  \tilAcal^{(b)*}_{\alpha'}(\lambda,p_z') D^{(-q_a)*}_{\beta'\alpha'}(\bx_\perp) \tilAcal^{(a)}_{\beta'}(\lambda,p_z')
 \bigg]
 \notag \\
 &= \sum_\lambda
 \frac{
  \mathcal A^{(a)}_\alpha(\lambda,p_z) \mathcal A^{(b)*}_\beta(\lambda,p_z)
 }{E - E^{(q_a)}_\lambda(p_z) + \rmi 0^+}
 \mp
 \sum_\lambda
 \frac{
  \tilde{\mathcal A}^{(b)*}_\beta(\lambda,-p_z) \tilde{\mathcal A}^{(a)}_\alpha(\lambda,-p_z)
 }{E + E^{(-q_a)}_\lambda(-p_z) + \rmi 0^+},
 \label{eq:retarded-spectral}
\end{align}
where we used the complex conjugation relation~\eqref{eq:D-complex-conjugation} for the second term and the orthogonality relation~\eqref{eq:D-orthogonality}.

Likewise, the time-ordered Feynman Green's function 
\begin{equation}
 \bar G^F_{ab}(x)
 :=
 -\rmi
 \bra{\Omega}
 \mathrm{T}[\hUcal^\dag(x) \hOcal_a(0) \hUcal(x) \, \hOcal_b^\dag(0)]
 \ket{\Omega}
\end{equation}
is shown to have the analogous spectral representation
\begin{equation}
 \bar G^{(ab),F}_{\alpha\beta}(E,p_z)
 =
 \sum_\lambda
 \frac{
  \mathcal A^{(a)}_\alpha(\lambda,p_z) \mathcal A^{(b)*}_\beta(\lambda,p_z)
 }{E - E^{(q_a)}_\lambda(p_z) + \rmi 0^+}
 \mp
 \sum_\lambda
 \frac{
  \tilAcal^{(b)*}_\beta(\lambda,-p_z) \tilAcal^{(a)}_\alpha(\lambda,-p_z)
 }{E + E^{(-q_a)}_\lambda(-p_z) - \rmi 0^+},
 \label{eq:time-ordered-spectral}
\end{equation}
which differs from the retarded representation only in the $\rmi 0^+$ prescription of the second pole.
The Euclidean Green's function follows by analytic continuation.

Equations~\eqref{eq:retarded-spectral} and~\eqref{eq:time-ordered-spectral} correspond to the familiar spectral representations for the retarded and time-ordered correlators.
Both representations share the same pole positions in the real-$E$ axis:
the first term has poles at $E = E^{(q_a)}_\lambda(p_z)$ corresponding to states in the charge-$q_a$ sector, while the second term has poles at $E = -E^{(-q_a)}_\lambda(-p_z)$ corresponding to states in the charge-$(-q_a)$ sector.
As usual, the two representations differ only in the $\rmi 0^+$ prescription of the poles.
Thus, all two-point Green's functions in the magnetic field are reconstructed from the spectral data of both charge sectors, $\{E^{(q_c)}_\lambda(p_z), \Acal^{(c)}_\alpha\}$ and $\{E^{(-q_c)}_\lambda(p_z), \tilAcal^{(c)}_\alpha\}$, together with the universal representation matrices $D^{(q)}_{\alpha\beta}(\bx_\perp)$ of the magnetic translation algebra.

\section{Summary and outlook}
\label{sec:summary}

In this paper, we have investigated quantum field theories in a static and uniform magnetic field from the viewpoint of their genuine global symmetries: magnetic translations and magnetic rotation.
Working on general grounds, without relying on specific models or perturbative expansions, we formulated these magnetic symmetries, derived their centrally-extended symmetry algebra [Eqs.~\eqref{eq:KK-commutator} and \eqref{eq:JK-commutator}] and its gauge-covariant realization, and clarified their consequences for correlation functions.
While the projective nature of magnetic translations is itself well known~\cite{Brown1964,Zak1964-1,Zak1964-2}, our emphasis is on its direct consequences for the structure of correlation functions in generic interacting theories, derived purely from symmetry.

We showed that the projective composition law~\eqref{eq:projective-rep.} of magnetic translation symmetry forces correlation functions of charged operators to acquire the Schwinger phase, leading to the factorized form~\eqref{eq:2-point-factorized-form} into a gauge-covariant phase factor~\eqref{eq:Phi-definition} and a reduced correlator depending only on relative coordinates.
In contrast to the conventional derivation based on explicit propagators in the proper-time representation, this identifies the Schwinger phase as a model-independent consequence of magnetic translation symmetry, valid non-perturbatively whenever the symmetry remains unbroken.
We further derived the selection rules imposed by the $\U(1)$ and magnetic rotation symmetries, and extended the analysis to higher-point functions, whose phase structure factorizes as in Eq.~\eqref{eq:N-point-factorized} and admits a natural geometric interpretation in terms of magnetic fluxes through coordinate-space triangles~\eqref{eq:N-point-flux}, discussed in, e.g., Ref.~\cite{Hattori:2023egw}.
Our derivation places this polygon phase on a purely symmetry-based footing.

Beyond these general constraints on correlation functions, we derived a symmetry-adapted spectral representation organized by representations of the magnetic translation algebra, which constitutes the main new result of this work.
Specifically, we derived the spectral representation~\eqref{eq:magnetic-spectral} together with its projected form~\eqref{eq:projected-green-spectral}.
Because the transverse magnetic translation generators satisfy the centrally-extended algebra~\eqref{eq:KK-commutator}, the ordinary transverse plane wave $\rme^{\rmi \bk_{\perp} \cdot \bx_{\perp}}$ no longer provides the natural organizing basis.
Instead, the transverse coordinate dependence is encoded in the representation matrices $D^{(q)}_{\alpha\beta}(\bx_\perp)$ of the magnetic translation algebra satisfying the projective composition law~\eqref{eq:D-projective}.

We then used this representation basis to derive a symmetry-adapted counterpart of the standard Umezawa--Kamefuchi--K\"all\'en--Lehmann spectral representation~\cite{Umezawa:1951rp,Kallen:1952zz,Lehmann:1954xi} in the presence of magnetic translation symmetry.
Different familiar descriptions, such as those based on the Landau gauge~\eqref{eq:projected-wightman-Landau} or the symmetric gauge~\eqref{eq:projected-wightman-rotational}, arise simply as different choices of basis for the same irreducible representation, and we showed how the retarded and time-ordered Green's functions are reconstructed from the same spectral data; see Eqs.~\eqref{eq:retarded-spectral} and \eqref{eq:time-ordered-spectral}.
To our knowledge, such a symmetry-adapted spectral representation organized directly by the magnetic translation algebra has not been formulated previously.
Our analysis thus provides a unified and model-independent framework for understanding the structure of correlation functions in magnetic fields directly from symmetry principles, and clarifies which characteristic structures---including the Schwinger phase and the nontrivial transverse structure of spectral representations---are fixed solely by the magnetic symmetry algebra itself.

Several directions deserve further investigation.

First, it would be interesting to extend the present analysis to in-medium systems at finite density.
Because the magnetic translation algebra is centrally extended by the
$\U(1)$ charge, finite $\U(1)$ charge density may imply spontaneous
symmetry breaking of magnetic translation symmetry and the appearance of associated Nambu--Goldstone modes~\cite{Nambu:1961tp,Goldstone:1961eq,Goldstone:1962es}.
This situation may share some similarities with systems possessing Galilean symmetry, where spacetime symmetry generators are centrally extended by the $\U(1)$ charge.
However, as in systems with spontaneously broken Galilean symmetry, spontaneous breaking of spacetime symmetries does not necessarily guarantee the existence of propagating gapless Nambu--Goldstone modes~\cite{Low:2001bw,Watanabe:2013iia,Nicolis:2013sga,Hayata:2013vfa,Brauner:2014aha,Hidaka:2014fra}.
It would therefore be interesting to clarify the low-energy dynamics
associated with the breaking of magnetic symmetries and to investigate
how the generalized Nambu--Goldstone theorem is realized in such systems.

The framework developed here can also be generalized to equilibrium in-medium systems at finite temperature and/or density.
Since our analysis is formulated directly in terms of symmetry and correlation functions, it may provide useful constraints on non-perturbative properties of interacting quantum matter in magnetic fields.
For example, the symmetry-based factorization and resulting spectral representation developed in this work could help disentangle universal kinematic structures from model-dependent dynamical information.

Another direction is to investigate how the magnetic symmetries studied in the present paper are realized in real-time effective descriptions, such as hydrodynamics or kinetic theory.
In particular, the projective structure of magnetic translation symmetry may impose nontrivial constraints on collective excitations, response functions, and transport phenomena in strongly magnetized media.
We leave these problems for future investigation.

\section*{Acknowledgment}

The authors thank Daiki Miura for stimulating discussions that motivated this work.
The authors also thank Yoshimasa Hidaka for useful discussions. 
M.H. is supported by the Japan Society for the Promotion of Science (JSPS) KAKENHI Grants No. 23K25870, No. 25K01002, No. 25K07316, and No. 26H01407.
K.N is supported by JSPS Fellows Grant 25KJ0148 and Grants No. 25K17392.
This work was partially supported by the RIKEN iTHEMS, the Niigata University Quantum Research Center (NU-Q), and the COREnet project of RCNP at the University of Osaka.

%\bibliographystyle{ptephy}
%\bibliography{sample}
%
% once the .bbl file has been generated then place the text in your article.

\appendix

\bibliographystyle{utphys}
\bibliography{refs}
\end{document}